\documentclass{pasj01}

\usepackage{comment}
\usepackage{url}
\usepackage{color}

\Received{$\langle$reception date$\rangle$}
\Accepted{$\langle$acception date$\rangle$}
\Published{$\langle$publication date$\rangle$}
\SetRunningHead{Astronomical Society of Japan}{Usage of \texttt{pasj00.cls}}

\begin{document}
\title{CO Multi-line Imaging of Nearby Galaxies (COMING). I\hspace{-0.02em}I\hspace{-0.02em}I. Dynamical effect on molecular gas density and star formation in the barred spiral galaxy NGC\,4303}
\author{
Yoshiyuki \textsc{Yajima}\altaffilmark{1,*},
Kazuo \textsc{Sorai}\altaffilmark{1,2,3,4,5},
Nario \textsc{Kuno}\altaffilmark{3,4},
Kazuyuki \textsc{Muraoka}\altaffilmark{6},
Yusuke \textsc{Miyamoto}\altaffilmark{7},
Hiroyuki \textsc{Kaneko}\altaffilmark{8},
Hiroyuki \textsc{Nakanishi}\altaffilmark{9},
Naomasa \textsc{Nakai}\altaffilmark{10},
Takahiro \textsc{Tanaka}\altaffilmark{3},
Yuya \textsc{Sato}\altaffilmark{3},
Dragan \textsc{Salak}\altaffilmark{10},
Kana \textsc{Morokuma-Matsui}\altaffilmark{11},
Naoko \textsc{Matsumoto}\altaffilmark{12},
Hsi-An \textsc{Pan}\altaffilmark{13},
Yuto \textsc{Noma}\altaffilmark{10},
Tsutomu T. \textsc{Takeuchi}\altaffilmark{14},
Moe \textsc{Yoda}\altaffilmark{14},
Mayu \textsc{Kuroda}\altaffilmark{6},
Atsushi \textsc{Yasuda}\altaffilmark{3},
Nagisa \textsc{Oi}\altaffilmark{15},
Shugo \textsc{Shibata}\altaffilmark{1},
Masumichi \textsc{Seta}\altaffilmark{10},
Yoshimasa \textsc{Watanabe}\altaffilmark{3,4},
Shoichiro \textsc{Kita}\altaffilmark{3},
Ryusei \textsc{Komatsuzaki}\altaffilmark{3},
Ayumi \textsc{Kajikawa}\altaffilmark{5},
and Yu \textsc{Yashima}\altaffilmark{5}
}

\altaffiltext{1}{Department of Cosmosciences, Graduate School of Science, Hokkaido University, Kita 10, Nishi 8, Kita-ku, Sapporo, Hokkaido 060-0810, Japan}
\altaffiltext{2}{Department of Physics, Faculty of Science, Hokkaido University, Kita 10, Nishi 8, Kita-ku, Sapporo, Hokkaido 060-0810, Japan}
\altaffiltext{3}{Graduate School of Pure and Applied Sciences, University of Tsukuba, 1-1-1 Tennodai, Tsukuba, Ibaraki 305-8571, Japan}
\altaffiltext{4}{Tomonaga Center for the History of the Universe, University of Tsukuba, 1-1-1 Tennodai, Tsukuba, Ibaraki 305-8571, Japan}
\altaffiltext{5}{Department of Physics, School of Science, Hokkaido University, Kita 10, Nishi 8, Kita-ku, Sapporo, Hokkaido 060-0810, Japan}
\altaffiltext{6}{Department of Physical Science, Osaka Prefecture University, 1-1 Gakuen, Sakai, Osaka 599-8531, Japan}
\altaffiltext{7}{National Astronomical Observatory of Japan, 2-21-1 Osawa, Mitaka, Tokyo 181-8588, Japan}
\altaffiltext{8}{Nobeyama Radio Observatory, 462-2 Nobeyama, Minamimaki, Minamisaku, Nagano 384-1305, Japan}
\altaffiltext{9}{Graduate School of Science and Engineering, Kagoshima University, 1-21-35 Korimoto, Kagoshima, Kagoshima 890-0065, Japan}
\altaffiltext{10}{Department of Physics, School of Science and Technology, Kwansei Gakuin University, 2-1 Gakuen, Sanda, Hyogo 669-1337, Japan}
\altaffiltext{11}{Institute of Space and Astronautical Science, Japan Aerospace Exploration Agency, 3-1-1 Yoshinodai, Chuo-ku, Sagamihara, Kanagawa 252-5210, Japan}
\altaffiltext{12}{The Research Institute for Time Studies, Yamaguchi University, 1677-1 Yoshida, Yamaguchi, Yamaguchi 753-8511, Japan}
\altaffiltext{13}{Academia Sinica, Institute of Astronomy and Astrophysics, No.1, Sec. 4, Roosevelt Rd, Taipei 10617, Taiwan}
\altaffiltext{14}{Division of Particle and Astrophysical Science, Nagoya University, Furo-cho, Chikusa-ku, Nagoya, Aichi 464-8602, Japan}
\altaffiltext{15}{Tokyo University of Science, Faculty of Science Division II, Liberal Arts, 1-3 Kagrazaka, Shinjuku-ku, Tokyo 162-8601, Japan}
\email{yajima@astro1.sci.hokudai.ac.jp}
\maketitle

\begin{abstract}
We present the results of \atom{C}{}{12}\atom{O}{}{}($J$\,=\,1--0) and \atom{C}{}{13}\atom{O}{}{}($J$\,=\,1--0) simultaneous mappings toward the nearby barred spiral galaxy NGC\,4303 as a part of the CO Multi-line Imaging of Nearby Galaxies (COMING) project.
Barred spiral galaxies often show lower star-formation efficiency (SFE) in their bar region compared to the spiral arms.
In this paper, we examine the relation between the SFEs and the volume densities of molecular gas $n(\mathrm{H_2})$ in the eight different regions within the galactic disk with \atom{C}{}{}\atom{O}{}{} data combined with archival far-ultraviolet and 24$\>\micron$ data.
We confirmed that SFE in the bar region is lower by 39\% than that in the spiral arms.
Moreover, velocity-alignment stacking analysis was performed for the spectra in the individual regions.
The integrated intensity ratios of \atom{C}{}{12}\atom{O}{}{} to \atom{C}{}{13}\atom{O}{}{} ($R_{12/13}$) range from 10 to 17 as the results of stacking.
Fixing a kinetic temperature of molecular gas, $n(\rm{H_2})$ was derived from $R_{12/13}$ via non-local thermodynamic equilibrium (non-LTE) analysis.
The density $n(\mathrm{H_2})$ in the bar is lower by 31--37\% than that in the arms and there is a rather tight positive correlation between SFEs and $n(\mathrm{H_2})$, with a correlation coefficient of $\sim$ 0.8.
Furthermore, we found a dependence of $n(\rm{H}_2)$ on the velocity dispersion of inter-molecular clouds ($\Delta V/\sin i$).
Specifically, $n(\mathrm{H_2})$ increases as $\Delta V/\sin i$ increases when $\Delta V/\sin i < 100$ km s$^{-1}$.
On the other hand, $n(\mathrm{H_2})$ decreases as $\Delta V/\sin i$ increases when $\Delta V/\sin i > 100$ km s$^{-1}$.
These relations indicate that the variations of SFE could be caused by the volume densities of molecular gas, and {the volume densities could be governed by the dynamical influence} such as cloud-cloud collisions, shear and enhanced inner-cloud turbulence.
\end{abstract}

\section{Introduction}
In the local universe, most of baryons in galaxies exist as stars and they are made from cold molecular gas.
That is, studying molecular gas, which is a raw material of the main component of galaxies is very important to understand galaxy evolution and formation.
Star-formation activity in barred spiral galaxies is often significantly different between in spiral arms and in their bar.
For example, previous studies have revealed that there are little new stars in the bar, while young stars are mainly distributed in spiral arms (e.g., \cite{Sheth00}, \yearcite{Sheth02}, \cite{Koda06}).
In other words, star-formation rate (SFR) is lower in the bar than that in arms.
Star-formation efficiency (SFE) is defined as the ratio of SFR to gas mass.
In barred spiral galaxies, SFE is also lower in the bar (\cite{Handa91}, \cite{Momose10}, \cite{Watanabe11}, \cite{Hirota14}).

\citet{Sorai12} suggested that most of molecular clouds in the offset ridges at leading sides of the bar are not gravitationally bound.
They concluded that vigorous gas motion in the bar ridges makes molecular gas unbound and SFE in the bar decreases since it is hard for molecular clouds to shrink and form stars; molecular clouds in the bar ridges may exist as diffuse gas.

Several studies have tried to reveal the relation between physical properties of molecular gas and star formation.
The \atom{H}{}{}\atom{C}{}{}\atom{N}{}{} molecule can trace high dense molecular gas because of its large electric dipole moment.
Therefore, integrated intensity ratio of \atom{H}{}{}\atom{C}{}{}\atom{N}{}{} and \atom{C}{}{12}\atom{O}{}{} emission lines can be regarded as a dense gas fraction of molecular gas.
There are several studies which have revealed that dense gas fraction has a positive correlation with both SFR and SFE (e.g., \cite{Solomon92}, \cite{Gao04}, \cite{Gao07}, \cite{Muraoka09}, \cite{Usero15}, \cite{Bigiel16}).
According to these results, it seems that dense gas fraction determines the activity of star formation.
Hence, it is important to understand how dense or diffuse molecular gas is distributed within a galaxy, i.e., volume density distribution.

However, deriving volume density of molecular gas (hereafter, we phrase ``density'' as ``volume density'') is not easy.
It cannot be estimated from only one emission line because one intensive variable such as an intensity of a line cannot constrain extensive variables such as density and temperature of molecular gas.
Thus, it is necessary to observe multiple lines
Unfortunately, since molecular emission lines excluding low-excitation line of \atom{C}{}{12}\atom{O}{}{} are usually weak, this confines observation targets to luminous (e.g., starburst galaxies; \cite{Iono07}, \cite{Pan15}) or very near galaxies (e.g., galaxies in the Local Group; \cite{Minamidani11}, \cite{Muraoka12}).
Therefore, molecular gas density in the bar of barred spiral galaxies is not yet clear.

A newly developed instrument enables us to observe weak lines with bright \atom{C}{}{}\atom{O}{}{} line simultaneously.
CO Multi-line Imaging of Nearby Galaxies (COMING) project\footnote{Official homepage is $\langle$\url{https://astro3.sci.hokudai.ac.jp/~radio/coming/index.html}$\rangle$} (Sorai et al. 2019), one of the legacy projects of Nobeyama Radio Observatory (NRO) produced \atom{C}{}{12}\atom{O}{}{}($J$\,=\,1--0), \atom{C}{}{13}\atom{O}{}{}($J$\,=\,1--0) and \atom{C}{}{}\atom{O}{}{18}($J$\,=\,1--0) simultaneous mappings toward a hundred and several tens of galaxies with the Nobeyama 45-m telescope and the multi-beam receiver FOur-beam REceiver System on 45-m Telescope (FOREST: \cite{Minamidani16}).

\citet{Muraoka16b} investigated physical properties of molecular gas in the barred spiral galaxy NGC\,2903.
They found a positive correlation between SFE and $n(\mathrm{H_2})$ among the regions.
As one of the papers from the COMING project, this paper presents the decreased SFEs in the bar caused by a lower density condition of molecular gas as shown in \citet{Muraoka16b}.
In addition, we also discuss the causes of SFE and $n(\mathrm{H_2})$ variations within a barred spiral galaxy as an extended study.

\begin{figure*}[t!]
 \begin{center}
  \includegraphics[width=15.5cm]{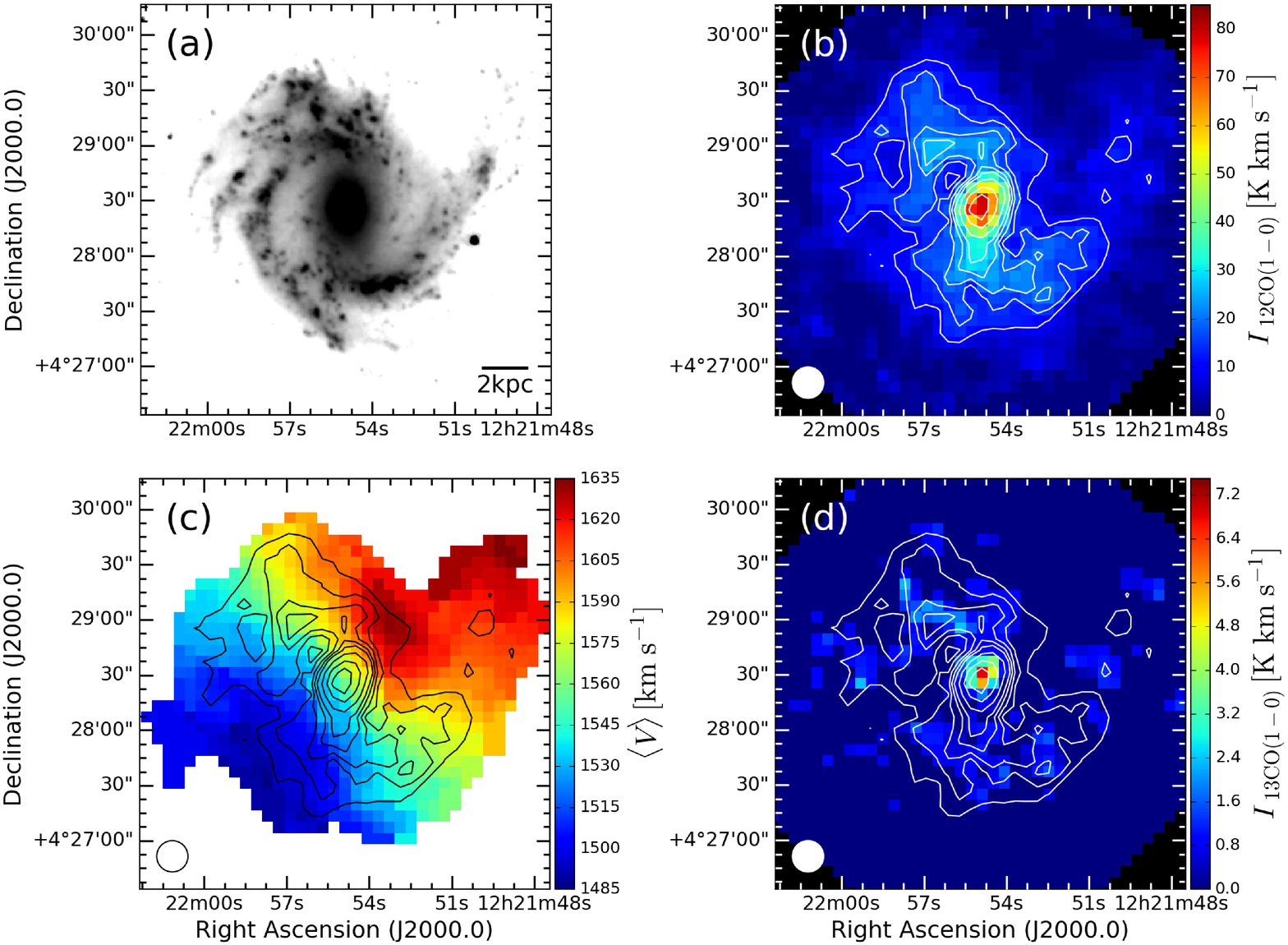}
 \end{center}
 \caption{(a) Spitzer/IRAC 3.6$\>\micron$ image of NGC\,4303 from S$^4$G survey \citep{Sheth10}.
 (b) The velocity-integrated intensity map of \atom{C}{}{12}\atom{O}{}{}($J$\,=\,1--0) line.
 Contour levels are 10, 15, 20, 25, 30, 40, 55, 70$\>$K$\>$km$\>$s$^{-1}$.
 The map-averaged 1$\,\sigma$ error of integrated intensities is 1.3$\>$K$\>$km$\>$s$^{-1}$.
 The white circle in the lower left corner represents the beam size of the Nobeyama 45-m telescope.
 (c) The intensity-weighted mean velocity field (color) measured by \atom{C}{}{12}\atom{O}{}{}($J$\,=\,1--0) line overlaid on \atom{C}{}{12}\atom{O}{}{}($J$\,=\,1--0) integrated intensities (contours). Contours are the same as (b).
 To get accurate velocity field map, we masked channels whose intensity is lower than 4\,$\sigma$.
 (d) The velocity-integrated intensity map of \atom{C}{}{13}\atom{O}{}{}($J$\,=\,1--0) line overlaid on \atom{C}{}{12}\atom{O}{}{}($J$\,=\,1--0) integrated intensities (contour). Contours are the same as (b).}
 \label{fig:Fig1}
\end{figure*}

To investigate the density of molecular gas in the bar, we study the barred spiral galaxy NGC\,4303 (M\,61) which is one of the targets in the COMING project.
Figure \ref{fig:Fig1} (a) shows the Spitzer/IRAC 3.6$\>\micron$ image obtained by \citet{Sheth10}.
This galaxy has three spiral arms (northern, eastern, and western), the bar and the bright nucleus.
The northern arm exists roughly from $(\alpha, \delta)_{\rm{J2000.0}} = (\timeform{12h21m54s}, \timeform{4D29'})$ to $(\timeform{12h21m57s}, \timeform{4D29'30"})$ and the eastern one is from $(\timeform{12h21m56s}, \timeform{4D29'})$ to $(\timeform{12h21m56s}, \timeform{4D27'24"})$ via $(\timeform{12h21m58s}, \timeform{4D28'20"})$ (see also figure \ref{fig:Fig3}).
NGC\,4303 is considered as a member of the Virgo cluster.
Because this galaxy is located in the outer region of the cluster, H\,\emissiontype{I} depletion is not seen in this galaxy (\cite{Cayatte90}). 
NGC\,4303 can be observed in the face-on view (the inclination angle is \timeform{27D}) and previous surveys showed that there is plenty of molecular gas in its bar (e.g., \cite{Kuno07}).
Hence, we can clearly recognize the bar and it is not difficult to analyze molecular gas in the bar.
Moreover, it is already confirmed in NGC\,4303 that SFE in the bar is lower than that in arms by \citet{Momose10}.
Therefore, this galaxy is one of the best objects to observe for discussing differences of molecular gas properties among the galactic structures.
Basic parameters of NGC 4303 are listed in table \ref{tab:Table1}.

\begin{table}[hbt]
  \begin{center}
  \tbl{Basic parameters of NGC\,4303.}{
  \begin{tabular}{lc} \hline
    Morphological type$^{*}$ & SAB(rs)bc \\
      Map center$^{\dag}$ & \\
      $\, \,$ Right ascension (J2000.0) & \timeform{12h21m54.895s} \\
      $\, \,$ Declination (J2000.0) & \timeform{4D28'25.13"} \\
      Distance$^{\ddag}$ & 16.5$\>$Mpc \\
      ${V_{\mathrm{LSR}}}^{*}$ & 1563$\>$km$\>$s$^{-1}$ \\
      Inclination angle$^{\S}$ & \timeform{27.0D} \\
      Position angle$^{\S}$ & \timeform{-36.4D} \\
      Linear scale & 80.0$\>$pc$\> \mathrm{arcsec^{-1}}$ \\
      Nucleus type$^{\|}$ & H$\,$\emissiontype{II}, Sy2 \\ \hline
  \end{tabular}}
  \label{tab:Table1}
  \begin{tabnote}
    $^{*}$ RC3.\\
    $^{\dag}$ \citet{Argyle90}.\\
    $^{\ddag}$ \citet{Mei07}.\\
    $^{\S}$ \citet{Salo15}.\\
    $^{\|}$ \citet{Filippenko85c}, \citet{Ho97d}.
  \end{tabnote}
  \end{center}
\end{table}

\section{Observations and data reduction}
We made simultaneous mappings of \atom{C}{}{12}\atom{O}{}{}($J$\,=\,1--0), \atom{C}{}{13}\atom{O}{}{}($J$\,=\,1--0) and \atom{C}{}{}\atom{O}{}{18}($J$\,=\,1--0) lines toward NGC\,4303 on 2016 February 12 and 13 with the Nobeyama 45-m radio telescope as a part of the COMING project.
The multi-beam receiver FOREST was used as the receiver front-end; it is a four-beam dual polarization receiver with sideband separation mixers.
The beam size is \timeform{15.0"} and \timeform{14.0"} at 110$\>$GHz and 115$\>$GHz, respectively, and the beam separation is \timeform{50"}. 

We adopted the On-The-Fly (OTF) mapping mode and used two scan patterns parallel to the major axis (X scan) and the minor axis (Y scan) of this galaxy, where the position angle of the X scan was set as $\timeform{-48D}$ against the north.
The scan separation was \timeform{4.975"}, and the scan speeds were $\timeform{15.1"}\> \rm{s}^{-1}$ for the X scan and $\timeform{14.8"}\> \rm{s}^{-1}$ for the Y scan.
The mapping area was \timeform{5.3'} $\times$ \timeform{5.1'}, which corresponds to $27 \times 25\>$kpc in this galaxy at the adopted distance and the inclination angle of this galaxy.
The receiver backend we used was SAM45 (\cite{Kuno11}, \cite{Kamazaki12}).
The total bandwidth and frequency resolution were 2$\>$GHz and 488$\>$kHz, respectively, which corresponds to 5330$\>$km$\>$s$^{-1}$ and 1.3$\>$km$\>$s$^{-1}$ at 115$\>$GHz.

The line intensity was calibrated by the chopper wheel method, yielding an antenna temperature, $T_{\rm{A}}^*$, corrected both atmospheric and antenna ohmic losses.
The main beam brightness temperature, $T_{\rm{MB}}$, was converted from $T_{\rm{A}}^*$, by observing a standard source, the carbon star IRC$\,+$10216.
The telescope pointing was checked and corrected every 30--50 minutes by observing the quasar 3C\,273, and the resultant pointing error was approximately \timeform{3"}.
The system noise temperature in $T_{\rm{A}}^*$ was 280--380$\>$K at 115$\>$GHz and 140--190$\>$K at 110$\>$GHz during the observations.

We used the auto-reduction system COMING ART (Sorai et al. 2019) for data reduction including removal of bad data, binning up of channels, basket-weaving \citep{Emerson88}, regridding and subtracting the baseline with the first- and third-degree polynomials.
This system was developed by members of the COMING project and based on the OTF data analysis package NOSTAR \citep{Sawada08} distributed by NRO.
The final velocity resolution and pixel scale were set to be 10$\>\ $km$\>$s$^{-1}$ and \timeform{6.0"}, respectively.
An effective angular resolution after map-smoothing is \timeform{17"} for all three lines, corresponding to 1.4$\>$kpc in this galaxy.
The root mean square (R.M.S.) noise levels of the maps were $\Delta T_{\mathrm{MB}} = 73\>$mK, 37$\>$mK and 36$\>$mK for \atom{C}{}{12}\atom{O}{}{}, \atom{C}{}{13}\atom{O}{}{} and \atom{C}{}{}\atom{O}{}{18}, respectively.

\section{Results}
\subsection{CO emission lines}
Figure \ref{fig:Fig1} (b) shows the velocity-integrated intensity $I_{\atom{C}{}{12}\atom{O}{}{}} \equiv \int T_{\rm{MB},\atom{C}{}{12}\atom{O}{}{}} dv$ (0th moment) map of \atom{C}{}{12}\atom{O}{}{}($J$\,=\,1--0).
Comparing the IRAC 3.6$\> \micron$ image [figure \ref{fig:Fig1} (a)], the western arm is clearly seen in the \atom{C}{}{12}\atom{O}{}{} line, whereas the eastern and the northern arms are ambiguous.
Assuming the standard \atom{C}{}{}\atom{O}{}{}-to-\atom{H}{}{}$_2$ conversion factor $X_{\mathrm{CO}}=2.0 \times 10^{20}\ \mathrm{cm^{-2}}\ (\mathrm{K\ km\ s^{-1}})^{-1}$ \citep{Bolatto13}, the surface density of molecular gas mass $\Sigma_{\rm{mol}}$ is measured by the following equation:
\begin{eqnarray}
\left( \frac{\Sigma_{\mathrm{mol}}}{M_{\Sol}\ \mathrm{pc^{-2}}} \right) &=&
1.36 \times 3.20 \cos i \left( \frac{I_{\atom{C}{}{12}\atom{O}{}{}}}{\mathrm{K\ km\ s^{-1}}} \right) \nonumber \\
& \times & \left\{ \frac{X_{\mathrm{CO}}}{2.0 \times 10^{20}\ \mathrm{cm^{-2}\ (K\ km\ s^{-1})^{-1}}} \right\}
\end{eqnarray} 
where $i$ is the inclination angle of this galaxy and a factor 1.36 is contribution of helium mass in molecular gas.
The other factor 3.20 represents the coefficient for conversion of units from $\mathrm{cm^{-2}}$ to $M_{\Sol}\> \mathrm{pc}^{-2}$.
{The total molecular gas mass in the map is $(9.40 \pm 0.06) \times 10^{9}\, M_{\Sol}$.}
This is almost consistent with previous results, such as \citet{Kuno07} ($9.7 \times 10^{9}\, M_{\Sol}$) and \citet{Momose10} ($9.1 \times 10^{9}\, M_{\Sol}$) corrected with the same distance, $X_{\mathrm{CO}}$ and the inclination angle.
Figure \ref{fig:Fig1} (c) shows the intensity-weighted mean velocity field $\langle V \rangle$ measured with \atom{C}{}{12}\atom{O}{}{} by the following equation: $\langle V \rangle \equiv \int vT_{\rm{MB},\atom{C}{}{12}\atom{O}{}{}} dv/ \int T_{\rm{MB},\atom{C}{}{12}\atom{O}{}{}} dv$ (1st moment).
To measure $\langle V \rangle$ accurately, velocity channels whose intensity is lower than 4\,$\sigma$ are masked.

\begin{figure*}[t!]
 \begin{center}
  \includegraphics[width=16cm]{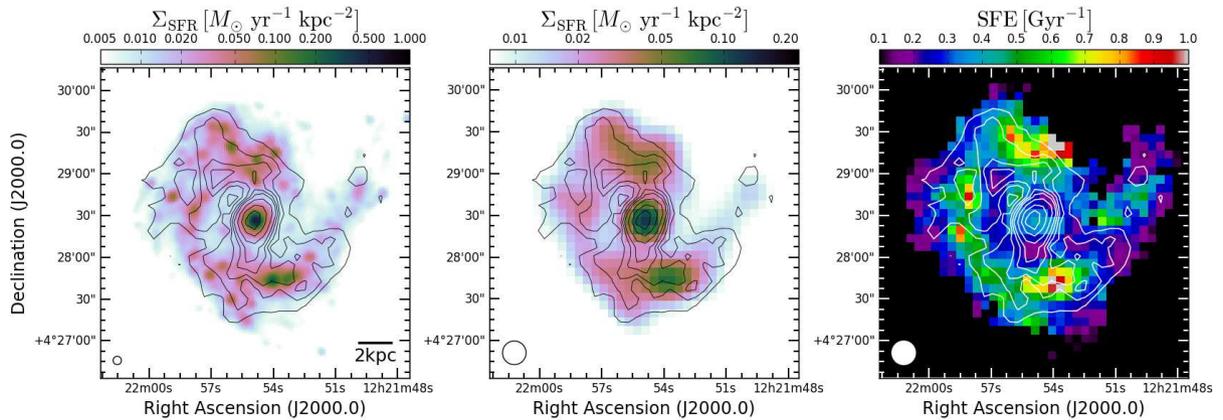}
 \end{center}
 \caption{(Left) Surface density of SFR in NGC\,4303 (color) overlaid on \atom{C}{}{12}\atom{O}{}{}($J$\,=\,1--0) integrated intensities (contours).
 The angular resolution of this panel is \timeform{6.0"} (the open circle in the bottom left corner) and the pixel scale is \timeform{1.5"} which are the same as the Spitzer/MIPS 24$\>\micron$ data.
 (Center) Convolved and regridded SFR surface density in NGC\,4303.
 The angular resolution and the grid spacing are the same as our CO data (\timeform{17"} and \timeform{6"}, respectively).
 The open circle in the bottom left corner indicates the angular resolution of this panel. 
 (Right) The SFE map (color) of NGC\,4303 overlaid on the \atom{C}{}{12}\atom{O}{}{}($J$\,=\,1--0) integrated intensity map (contours).
 Pixels whose integrated intensity of \atom{C}{}{12}\atom{O}{}{}($J$\,=\,1--0) line is below 3$\,\sigma$ were masked.
 The angular resolution (the open circle in the bottom left corner) and the grid spacing are the same as the middle panel (\timeform{17"} and \timeform{6"}, respectively).
 }
 \label{fig:Fig2}
\end{figure*}

Figure \ref{fig:Fig1} (d) shows the \atom{C}{}{13}\atom{O}{}{} integrated intensity map.
Because of low abundance of \atom{C}{}{13}, the \atom{C}{}{13}\atom{O}{}{} emission is optically thin and very weak.
Thus, the signal-to-noise ratio ($S/N$) is poor, although the R.M.S. noise level of \atom{C}{}{13}\atom{O}{}{} is about half of that of \atom{C}{}{12}\atom{O}{}{}.
Emission of \atom{C}{}{13}\atom{O}{}{} can be mainly seen from the galactic center.
We could not detect any emission of \atom{C}{}{}\atom{O}{}{18}($J$\,=\,1--0).

\subsection{Star-formation rate and efficiency}
From the combination of far ultraviolet (FUV) or H$\alpha$ line, and far infrared intensities, SFR can be derived (e.g., \cite{Calzetti07}, \cite{Kennicutt07}, \cite{Leroy08}).
In this paper, we adopted the following equation reported by \citet{Leroy08}:
\begin{eqnarray}
\Sigma_{\mathrm{SFR}}[ &M_{\Sol}& \> \mathrm{yr^{-1}\> kpc^{-2}}] =
( 8.1 \pm 0 \times 10^{-2} I_{\mathrm{FUV}}\> [\mathrm{MJy\> sr^{-1}}] \nonumber \\
&&\>\>\>\>\>\>\>\>\>\>\>\>\>\>\>\>\>\> +3.2^{+1.2}_{-0.7} \times 10^{-3} I_{\mathrm{24\micron}}\> [\mathrm{MJy\> sr^{-1}}]) \cos i
\end{eqnarray}
where $I_{\mathrm{FUV}}$ and $I_{\mathrm{24\micron}}$ are intensities of FUV and 24$\>\micron$ continuum, respectively.
This equation adopts Kroupa initial mass function (IMF, \cite{Kroupa01}) with mass range of 0.1--120$\> M_{\Sol}$.
Note that SFR assuming this IMF is less by $1/1.59$ than that derived by the Salpeter IMF \citep{Salpeter55} of 0.1--100\,$M_{\Sol}$ \citep{Leroy08}.

We made use of GALEX FUV \citep{Paz07} and Spitzer/MIPS 24 $\micron$ \citep{Bendo12} archival data downloaded from the NASA/IPAC Extragalactic Database\footnote{$\langle$\url{https://ned.ipac.caltech.edu}$\rangle$} (NED).
Since the original angular resolution of GALEX FUV and Spitzer/MIPS 24$\>\micron$ data are \timeform{4.0"} and \timeform{6.0"}, respectively, {the FUV and 24$\>\micron$ data were convolved to match Gaussian beams of \timeform{17"}, which is the same as our \atom{C}{}{}\atom{O}{}{} data.}
They were also regridded into \timeform{6.0"} to match our \atom{C}{}{}\atom{O}{}{} data because their pixel scale are \timeform{1.5"}. 
To exhibit lower $\Sigma_{\mathrm{SFR}}$ in the bar, we also made a $\Sigma_{\mathrm{SFR}}$ map whose resolution is \timeform{6.0"} and pixel scale is \timeform{1.5"} from convolved FUV and original 24$\>\micron$ image.
The left panel of figure \ref{fig:Fig2} shows it as a high resolution $\Sigma_{\rm{SFR}}$ map.
SFE can be calculated by the following equation:
\begin{equation}
\left( \frac{\mathrm{SFE}}{\mathrm{yr^{-1}}} \right) =
\left( \frac{\Sigma_{\mathrm{SFR}}}{M_{\Sol}\ \mathrm{yr^{-1}\ pc^{-2}}} \right) \bigg/ \left( \frac{\Sigma_{\mathrm{mol}}}{M_{\Sol}\ \mathrm{pc^{-2}}} \right).
\end{equation}
SFEs in each pixel were measured with our \atom{C}{}{12}\atom{O}{}{} data and the convolved and regridded SFR map (the middle panel of figure 2).
The right panel of figure \ref{fig:Fig2} shows the result.

In the higher resolution $\Sigma_{\mathrm{SFR}}$ map (the left panel of figure \ref{fig:Fig2}) there are little newborn stars in the bar, while many clumpy structures (H\,\emissiontype{II} regions) are found in bar ends and spiral arms.
The central region of NGC\,4303 also shows higher SFR.
In the right panel of figure \ref{fig:Fig2}, SFE was measured for pixels whose $I_{\atom{C}{}{12}\atom{O}{}{}}$ is larger than 3\,$\sigma$ error.
SFE is higher in spiral arms, especially the northern arm and the southern bar end.
On the other hand, SFE is depressed in the bar, outer disk and inter-arms region.

\subsection{Regionally averaged quantities in each galactic structure}
\begin{figure}[b!t!]
 \begin{center}
  \includegraphics[width=7.8cm]{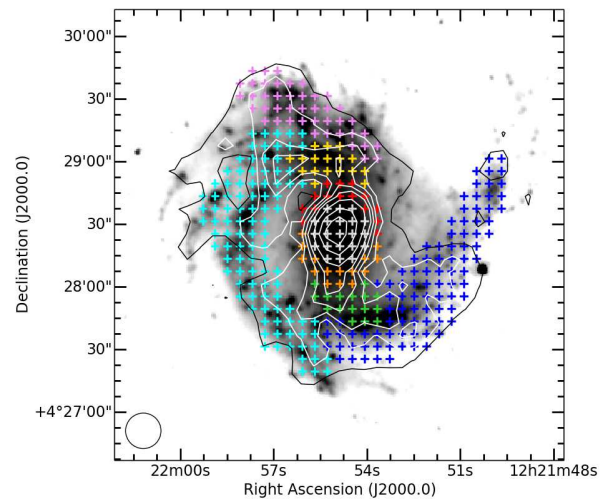}
 \end{center}
 \caption{The separated regions based on the galactic structures: the center (gray), the northern bar (red), the southern bar (orange), the northern bar end (yellow), the southern bar end (green), the northern arm (purple), the eastern arm (cyan) and the western arm (blue) overlaid on the Spitzer/IRAC 3.6$\>\micron$ image (gray scale) and \atom{C}{}{12}\atom{O}{}{}($J$\,=\,1--0) integrated intensities (contours). Contours are the same as figure \ref{fig:Fig1} (b).
              The open circle in the bottom left corner represents the beam size of the Nobeyama 45-m telescope \timeform{17"}.
              Each cross marker corresponds to pixels of the COMING original data (\timeform{6"} $\times$ \timeform{6"}).}
 \label{fig:Fig3}
\end{figure}

To compare physical quantities such as $\Sigma_{\mathrm{mol}}$, $\Sigma_{\mathrm{SFR}}$ and SFE among galactic structures, we separated NGC\,4303 into eight components: the center, northern bar, southern bar, northern bar end, southern bar end, northern arm, eastern arm and western arm according to figure \ref{fig:Fig3}.
To determine each region, we referred \atom{C}{}{12}\atom{O}{}{}($J$\,=\,1--0) contours and the near infrared image in figure \ref{fig:Fig1} (a).
We measured $\Sigma_{\mathrm{mol}}$, $\Sigma_{\mathrm{SFR}}$ and SFE in each region.
Regionally averaged $\Sigma_{\mathrm{SFR}}$ was measured from the convolved and regridded map to match our \atom{C}{}{}\atom{O}{}{} data (the middle panel of figure 2) and SFE was measured from regionally averaged $\Sigma_{\mathrm{SFR}}$ over $\Sigma_{\mathrm{mol}}$, not averaged SFE in each pixel.
Averaged $\Sigma_{\mathrm{mol}}$, $\Sigma_{\mathrm{SFR}}$ and SFE are listed in table \ref{tab:Table2}.

Even though the left panel of figure \ref{fig:Fig2} shows prominently lower $\Sigma_{\rm{SFR}}$ in the bar, regionally averaged values listed in table \ref{tab:Table2} are not so different compared with that in the arms.
This is because the convolution process smooths distribution of $\Sigma_{\mathrm{SFR}}$ and it closes the spatial gap within this galaxy (cf. the left and middle panel of figure 2).
The SFE in the bar is lower by approximately 39\% than that in the arms.
By contrast, SFE in bar ends is slightly higher (by $\sim 10\%$) than that in arms.
\citet{Momose10} also reported that the SFE in the bar of NGC\,4303 is lower by 43\% than that in the arms region.
Our result is almost consistent with \citet{Momose10} and {the difference would be originated from the definition of each region and the angular resolution (\timeform{17"} and \timeform{6"}).}

To derive density of molecular gas, we need integrated intensity ratios; the ratio of $I_{\atom{C}{}{12}\atom{O}{}{}}$ to $I_{\atom{C}{}{13}\atom{O}{}{}} \equiv R_{12/13}$ needs to be measured.
However, as shown in figure \ref{fig:Fig1} (d), \atom{C}{}{13}\atom{O}{}{}($J$\,=\,1--0) line is not strong enough to measure integrated intensities.
Therefore, we performed the stacking analysis (\cite{Schruba11}, \cite{Morokuma15}) in order to measure \atom{C}{}{13}\atom{O}{}{}($J$\,=\,1--0) integrated intensity $I_{\atom{C}{}{13}\atom{O}{}{}}$ accurately.
The process of the stacking analysis is as follows:
Firstly, spectra are shifted along the velocity axis according to intensity-weighted mean velocity field measured in \atom{C}{}{12}\atom{O}{}{}($J$\,=\,1--0) [figure \ref{fig:Fig1} (c)].
Secondly, these velocity-shifted spectra are averaged.
In this way, the noise level decreases and we can get a $S/N$-improved spectrum although spatial information is lost.
{Since the grid size of the data we used for the stacking analysis is \timeform{6"} (the same as COMING original data, see figure 3), the spectra are oversampled.}

\begin{figure*}[b!t!]
 \begin{center}
  \includegraphics[width=14.5cm]{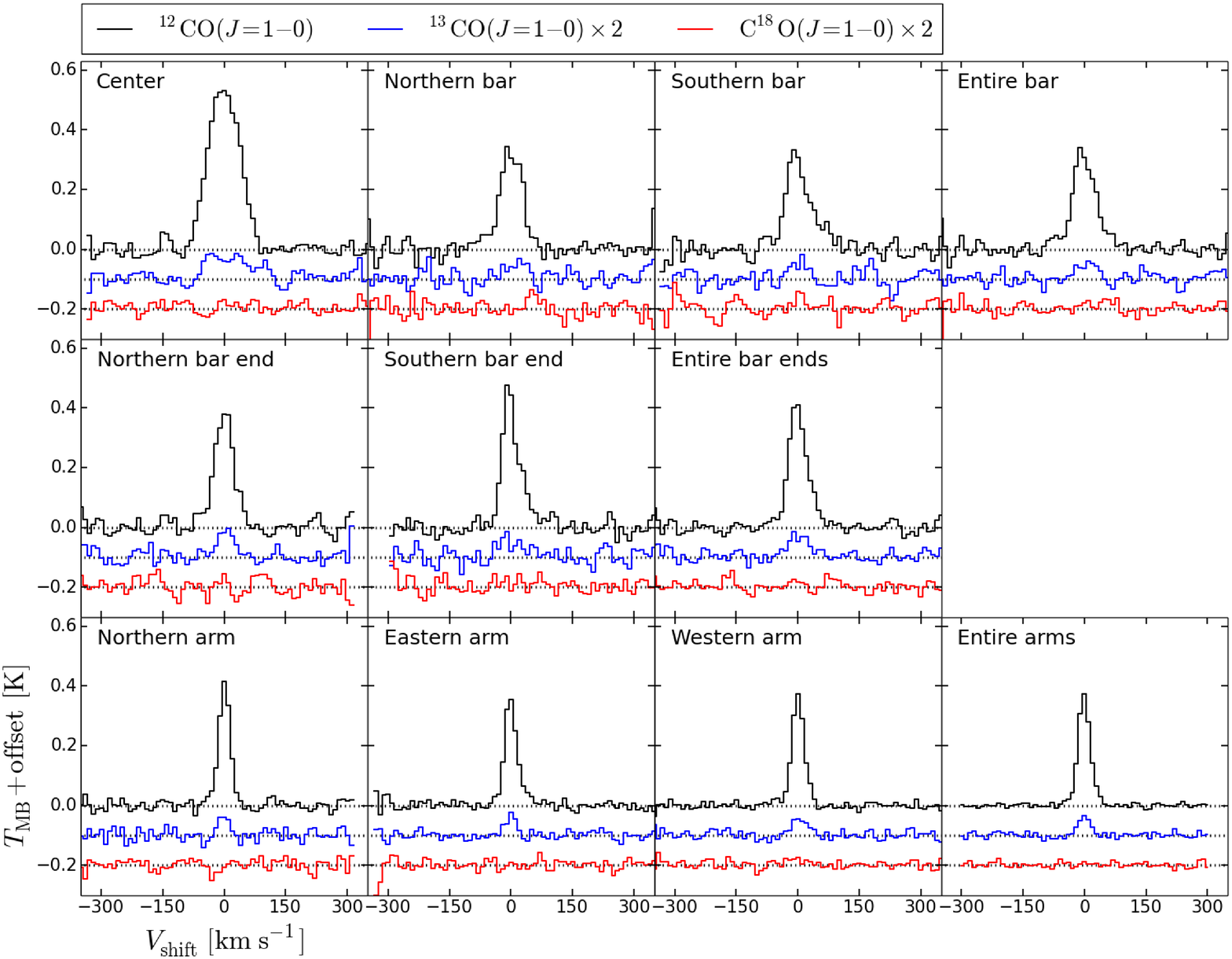}
 \end{center}
 \caption{Stacked \atom{C}{}{12}\atom{O}{}{}($J$\,=\,1--0) (black), \atom{C}{}{13}\atom{O}{}{}($J$\,=\,1--0) (blue) and \atom{C}{}{}\atom{O}{}{18}($J$\,=\,1--0) (red) spectra in each region. Offsets are 0, -0.1, -0.2 K for \atom{C}{}{12}\atom{O}{}{}($J$\,=\,1--0), \atom{C}{}{13}\atom{O}{}{}($J$\,=\,1--0) and \atom{C}{}{}\atom{O}{}{18}($J$\,=\,1--0), respectively. Intensities of \atom{C}{}{13}\atom{O}{}{}($J$\,=\,1--0) and \atom{C}{}{}\atom{O}{}{18}($J$\,=\,1--0) spectra are multiplied by two.}
 \label{fig:Fig4}
\end{figure*}

Figure \ref{fig:Fig4} shows stacked \atom{C}{}{12}\atom{O}{}{}($J$\,=\,1--0), \atom{C}{}{13}\atom{O}{}{}($J$\,=\,1--0) and \atom{C}{}{}\atom{O}{}{18}($J$\,=\,1--0) spectra in each region.
Parameters of stacked spectra are also summarized in Table \ref{tab:Table2}.
We measured integrated intensities for \atom{C}{}{12}\atom{O}{}{}($J$\,=\,1--0) and \atom{C}{}{13}\atom{O}{}{}($J$\,=\,1--0) stacked spectra, which is significantly detected.
Note that integrated intensities in some regions such as the entire bar are not necessarily average of values in its finer regions such as the northern and southern bars.
This is because integrated intensities can be changed if $S/N$ of stacked spectra is different and it depends on how spectra are stacked.
Thanks to the stacking analysis, \atom{C}{}{13}\atom{O}{}{}($J$\,=\,1--0) emission could be detected in all regions.
On the other hand, \atom{C}{}{}\atom{O}{}{18}($J$\,=\,1--0) emission line could not be detected.

The number of stacked spectra is approximately 15 for each part of the bar, the bar ends and the center, 40 for the northern arm and 90 for the eastern and the western arm (figure 3).
Correspondingly, the noise level after stacking in the above-mentioned regions is typically 20, 15, 10$\>$mK for \atom{C}{}{12}\atom{O}{}{}, and 11,  9, 5$\>$mK for \atom{C}{}{13}\atom{O}{}{}, respectively.
The resultant noise levels of stacked spectra are approximately larger by a factor of $\sim 1.2$ than expected (the inverse number of square root of the number of the stacked spectra).
This would be due to oversampling of the spectra, i.e., the noises for adjacent spectra are not random because of oversampling.

While several studies have reported that $R_{12/13}$ is high in starburst galaxies, luminous and ultra luminous infrared galaxies (LIRG, ULIRG; e.g., \cite{Aalto91} and \cite{Aalto95}), it ranges 5--18 for normal spiral galaxies (\cite{Paglione01}, \cite{Cao17}, \cite{Cormier18}, Y. Sato et al. in preparation).
Measured $R_{12/13}$ from stacked spectra in each region of NGC\,4303 are consistent with such normal galaxies.
Furthermore, $R_{12/13}$ is lower in the bar and rather higher in the bar end and the center where star formation is suppressed and enhanced, respectively.
This tendency is seen in another nearby barred spiral galaxy NGC\,3627 (\cite{Watanabe11}, \cite{Morokuma15}).
We discuss physical reasons for $R_{12/13}$ variations of details in the next section.

The 3$\sigma$ lower limits of the \atom{C}{}{13}\atom{O}{}{}/\atom{C}{}{}\atom{O}{}{18} integrated intensity ratios are also measured for the representative regions (the entire arms, the entire bar ends, the bar and the center).
They typically range from 2.2 (for the entire bar) to 4.4 (for the center). 
This result falls within the range of the values for other local spiral galaxies (e.g., \cite{JD17}).

In addition, the full width at half maximum (FWHM) of \atom{C}{}{12}\atom{O}{}{}($J$\,=\,1--0) line profiles $\Delta V$ was measured without Gaussian fitting.
We interpolated channels so that velocity whose intensity becomes the half of the peak was determined.  Among stacked spectra, we found that \atom{C}{}{12}\atom{O}{}{}($J$\,=\,1--0) spectra in spiral arms look sharp in profile, while they are rather broader in the bar. 
In fact, the $\Delta V$ is smaller in the spiral arm regions than the bar regions.  

\renewcommand\arraystretch{1.25}
\begin{table*}[b!t!]
  \begin{center}
  \tbl{Regionally-averaged quantities.}{
  \begin{tabular}{lccccccc} \hline
                               &$\Sigma_{\mathrm{mol}} $& $\Sigma_{\mathrm{SFR}}$                                   & SFE                                 &$I_{\atom{C}{}{12}\atom{O}{}{}}$ &$I_{\atom{C}{}{13}\atom{O}{}{}}$ &$R_{12/13}$     & $\Delta V$ \\
                               &[$M_{\Sol}\ \mathrm{pc}^{-2}$]      &[$10^{-2}\ M_{\Sol}\mathrm{\ yr^{-1}\ kpc^{-2}}$]&[$10^{-10}\ \mathrm{yr}^{-1}$]     & [$\mathrm{K\ km\ s^{-1}}$]                & [$\mathrm{K\ km\ s^{-1}}$]                 & &[$\mathrm{km\ s^{-1}}$] \\ \hline
    Center                 & $201 \pm 2.5$                & $6.49 ^{+ 0.57} _{-0.33}$                                     & $3.23 ^{+ 0.29} _{-0.17}$ & $51.2\pm1.0$                  & $4.5\pm0.3$                     & $11.4\pm0.8$ & $95\pm5$\\
    Northern bar        & $83.5 \pm 2.1$              & $2.38 ^{+ 0.23} _{-0.14}$                                      & $2.85 ^{+ 0.29} _{-0.18}$ & $22.6\pm0.9$                  & $1.4\pm0.3$                    & $16.6\pm3.4$ & $65\pm4$\\
    Southern bar       & $88.8 \pm 2.4$               & $2.09 ^{+ 0.20} _{-0.12}$                                      & $2.36 ^{+ 0.24} _{-0.15}$ & $22.9\pm0.9$                  & $1.4\pm0.2$                    & $16.4\pm2.8$ & $59\pm12$\\
    $\>\>\>\>$Entire bar             & $86.1 \pm 1.6$              & $2.23 ^{+ 0.16} _{-0.09}$                                      & $2.60 ^{+ 0.19} _{-0.12}$ & $23.5\pm0.7$                  & $1.4\pm0.2$                    & $17.0\pm2.3$ & $64\pm4$\\ 
    Northern bar end & $82.3 \pm 1.8$              & $3.51 ^{+ 0.33} _{-0.19}$                                       & $4.27 ^{+ 0.41} _{-0.25}$ & $23.0\pm0.8$                 & $2.1\pm0.2$                    & $10.8\pm1.3$ & $53\pm4$\\
    Southern bar end & $89.4 \pm 2.1$              & $4.46 ^{+ 0.45} _{-0.26}$                                      & $5.00 ^{+ 0.52} _{-0.32}$ & $23.1\pm0.6$                  & $2.3\pm0.4$                    & $10.2\pm1.9$ & $41\pm4$\\ 
    $\>\>\>\>$Entire bar ends    & $85.7 \pm 1.4$              & $3.97 ^{+ 0.28} _{-0.16}$                                       & $4.63 ^{+ 0.33} _{-0.20}$ & $22.6\pm0.5$                 & $2.1\pm0.2$                    & $10.7\pm1.1$ & $51\pm3$\\
    Northern arm       & $55.6 \pm 1.0$              & $2.97 ^{+ 0.18} _{-0.10}$                                       & $5.34 ^{+ 0.34} _{-0.21}$ & $14.7\pm0.4$                 & $1.1\pm0.2$                   & $13.7\pm2.6$ & $32\pm2$\\ 
    Eastern arm        & $53.4 \pm 0.6$               & $2.20 ^{+ 0.08} _{-0.05}$                                       & $4.13 ^{+ 0.17} _{-0.11}$ & $14.0\pm0.3$                 & $1.1\pm0.1$                   & $12.5\pm1.1$ & $35\pm2$\\
    Western arm       & $52.3 \pm 0.6$               & $2.03 ^{+ 0.09} _{-0.05}$                                       & $3.87 ^{+ 0.19} _{-0.11}$ & $14.4\pm0.3$                 & $1.1\pm0.1$                   & $13.2\pm1.6$ & $34\pm1$\\
    $\>\>\>\>$Entire arms         & $53.3 \pm 0.4$               & $2.27 ^{+ 0.06} _{-0.04}$                                      & $4.26 ^{+ 0.12} _{-0.07}$ & $14.1\pm0.2$                 & $1.1\pm0.1$                   & $12.5\pm1.1$  & $34\pm1$\\ 
    \hline
  \end{tabular}} 
  \label{tab:Table2}
  \end{center}
\end{table*}


\section{Discussion}
\subsection{{What $R_{12/13}$ physically means}}
If \atom{C}{}{12}\atom{O}{}{}($J$\,=\,1--0) emission is optically thick, \atom{C}{}{13}\atom{O}{}{}($J$\,=\,1--0) is optically thin and the LTE assumption is applicable, $R_{12/13}$ is proportional to ${\tau_{13}}^{-1}$, where $\tau_{13}$ is optical depth of \atom{C}{}{13}\atom{O}{}{}.
In addition, when excitation temperature of \atom{C}{}{13}\atom{O}{}{}, $T_{\rm{ex,13}}$ satisfies $T_{\mathrm{ex,13}} \gg 5.3\>$K [$\sim h \nu_{\atom{C}{}{13}\atom{O}{}{}} / k_{\rm{B}}$, where $h$ is the Planck constant, $\nu_{\atom{C}{}{13}\atom{O}{}{}}$ is the rest frequency of \atom{C}{}{13}\atom{O}{}{}($J$\,=\,1--0) and $k_{\rm{B}}$ is the Boltzmann constant], $\tau_{13}$ is approximately proportional to $\mathcal{N}_{\atom{C}{}{13}\atom{O}{}{}} / ({T_{\mathrm{kin}}}^2 dv)$, where $\mathcal{N}_{\atom{C}{}{13}\atom{O}{}{}}$ is column density of \atom{C}{}{13}\atom{O}{}{} and $dv$ is velocity dispersion within a molecular cloud. 
Therefore, some studies suggest that $R_{12/13}$ strongly depends on $T_{\rm{kin}}$ (e.g., \cite{Paglione01}, \cite{Hirota10}).
However, it could be doubtful that \atom{C}{}{13}\atom{O}{}{} satisfies LTE condition because of its low abundance ratio and low column density.
In this subsection, we discuss what changes $R_{12/13}$ focusing particularly on the chemical abundance ratio, velocity dispersion and physical properties of molecular gas such as $n({\rm{H_2}})$ and $T_{\rm{kin}}$.

When $T_{\rm{kin}} \lesssim 35\>$K, chemical fractionation (e.g., $\atom{C}{}{12}\atom{O}{}{} + \atom{C}{}{13}^+ \rightleftharpoons \atom{C}{}{13}\atom{O}{}{} + \atom{C}{}{12}^+$) is promoted (e.g., \cite{Langer84}, \cite{Liszt07}).
In addition, \atom{C}{}{13}\atom{O}{}{} can be selectively dissociated by radiation from massive stars (e.g., \cite{Davis14}).
On the other hand, \citet{Szucs14} showed there are no strong effects on the \atom{C}{}{13}\atom{O}{}{} abundance ratio by selective photodissociation.
\citet{Milam05} indicated both processes of chemical fractionation and selective photodissociation are inconsistent with the abundance ratios of [\atom{C}{}{12}]/[\atom{C}{}{13}] in the Milky Way.
\citet{Cao17} presumed the effects on $R_{12/13}$ by changes in the chemical abundance ratio are small enough to be easily washed out by other factors based on the discrepancy between the $R_{12/13}$ and the abundance ratios of [\atom{C}{}{12}]/[\atom{C}{}{13}] in the Galactic center and inner disk of the Milky Way.
Hence, we consider $R_{12/13}$ is mainly changed by other factors except for the abundance ratio.

Next, how can velocity dispersion influence $R_{12/13}$?
If velocity dispersion is large, $\tau_{13}$ becomes thin.
\citet{Watanabe11} reported the spatially resolved $R_{12/13}$ in the barred spiral galaxy NGC\,3627 and $R_{12/13}$ in the bar is less than in other regions such as the spiral arms.
They argue the depletion of $R_{12/13}$ would be caused by the optically thinner \atom{C}{}{13}\atom{O}{}{} due to high velocity dispersion in the bar.
By contrast, \citet{Meier04} showed there is no correlation between $R_{12/13}$ and the velocity dispersions of \atom{C}{}{13}\atom{O}{}{} spectra in the nucleus of NGC\,6946.
\citet{Cao17} also showed no correlations between $R_{12/13}$ and the intensity-weighted velocity dispersions of \atom{C}{}{12}\atom{O}{}{} spectra for 11 galaxies.
Note that their spatial resolution is different.
\citet{Watanabe11} performed a single-dish observation and the resolution is $\sim 1\>$kpc.
While the others were interferometric, thus the spatial resolution corresponds to sub-kpc or a few 100$\>$pc.
As seen above, the effect of velocity dispersion on $R_{12/13}$ is under debate.
Therefore, we need to calculate $\tau_{13}$ practically in a certain way.

At last, we examine the effects of physical properties such as density or temperature.
As mentioned in the first part of this subsection, if the four conditions (the optical depth of \atom{C}{}{12}\atom{O}{}{} $\tau_{12} \gg 1$, $\tau_{13} \ll 1$, the LTE approximation and $T_{\rm{ex,13}} \gg 5.3\>$K) are applicable, $R_{12/13}$ is proportional to ${T_{\rm{kin}}}^{-2}$.
However, if the LTE approximation is not applicable, the temperature dependence of $R_{12/13}$ becomes weak as suggested in \citet{Paglione01}.
Alternatively, \citet{Meier04} and \citet{Meier08} suggested that sub-thermally excited \atom{C}{}{13}\atom{O}{}{} is significant for variations of $R_{12/13}$.
The critical density of \atom{C}{}{13}\atom{O}{}{}($J$\,=\,1--0) is $\sim 2\times 10^3 \>$cm$^{-3}$, whereas that of \atom{C}{}{12}\atom{O}{}{}($J$\,=\,1--0) is $\sim 300\>$cm$^{-3}$.
Therefore, when $n(\rm{H_2})$ becomes less than $\sim 300\>$cm$^{-3}$, \atom{C}{}{13}\atom{O}{}{} becomes strongly sub-thermalized and the intensity of \atom{C}{}{13}\atom{O}{}{}($J$\,=\,1--0) is rapidly decreased.
That is, $R_{12/13}$ is sensitive to the change in $n(\rm{H_2})$.
Although one may presume that kinetic temperatures are higher in vicinities of star-forming regions, several studies reported the kinetic temperatures around it are not higher than other regions (e.g., \cite{Meier04}, \cite{Schinnerer10}).
Feedback from massive stars would have a small impact on kinetic temperature of molecular gas.
Hence, we suggest that $R_{12/13}$ can be a indicator of $n(\rm{H_2})$, as long as \atom{C}{}{13}\atom{O}{}{} is strongly sub-thermalized.

\subsection{Derivation of molecular gas density via non-LTE analysis}
With the non-local thermodynamic equilibrium (non-LTE) analysis, we can derive density and kinetic temperature of molecular gas from measured integrated intensity ratios by modeling observed emission lines.
It presumes internal motion of a molecular cloud with microscopic turbulences or a systematic acceleration field to calculate escape probability of photons, and computes intensities of emission lines.
The latter case is called the large velocity gradient (LVG) method (e.g., \cite{Scoville74}, \cite{Goldreich74}).
In this study, we use an open source code for the non-LTE analysis RADEX \citep{vanderTak07}\footnote{See also $\langle$\url{http://home.strw.leidenuniv.nl/~moldata/radex_manual.pdf}$\rangle$} based on the former case.

\subsubsection{{Settings of non-LTE analysis}}
RADEX solves the statistical equilibrium equations and the radiative transfer equation iteratively for an input molecule at a given kinetic temperature $T_{\mathrm{kin}}$ and a number density of molecular hydrogen $n(\atom{H}{}{}_2)$, and outputs a brightness temperature $T_{\mathrm{b}}$ and an integrated intensity $I$ of a single molecular cloud for each emission line of the molecule.
However, this computed $I$ cannot be compared with observed integrated intensities directly because observation results are affected by beam dilution.
This problem can be solved by taking a ratio of two integrated intensities in order to cancel the beam dilution effect with assumptions that the medium is uniform and the solid angle subtended by emissions of each line is the same.
In summary, RADEX computes integrated intensity ratios assuming $n(\rm{H_2})$ and $T_{\rm{kin}}$ of molecular gas.

RADEX can select three types of a cloud geometry (expanding or shrinking sphere, uniform sphere, plane slab) to calculate photon escape probability.
We adopted the expanding and shrinking model, and in that case it is calculated with the Sobolev approximation \citep{Sobolev60}.
Collision partners to \atom{C}{}{12}\atom{O}{}{} and \atom{C}{}{13}\atom{O}{}{} were set only to \atom{H}{}{}$_2$.
The brightness temperature of the background radiation was set to 2.73$\>$K.
To derive both of $n(\mathrm{H_2})$ and $T_{\mathrm{kin}}$ simultaneously, three emission lines, or two independent integrated intensity ratios are needed at least.
There are two available lines.
However, we have only one integrated intensity ratio $R_{12/13}$.
In this paper, RADEX is calculated with a fixed $T_{\mathrm{kin}}$ ($T_{\mathrm{kin}} = 10$, 15, 20$\>$K).
According to the previous subsection, this assumption is not invalid because $T_{\rm{kin}}$ would not exert a strong influence on $R_{12/13}$.

In RADEX calculations, a \atom{C}{}{12}\atom{O}{}{} column density $(\mathcal{N}_{\mathrm{\atom{C}{}{12}\atom{O}{}{}}})$, \atom{C}{}{13}\atom{O}{}{} column density $(\mathcal{N}_{\mathrm{\atom{C}{}{13}\atom{O}{}{}}})$ and FWHM of a cloud $(dv)$ are needed as input parameters.
$\mathcal{N}_{\mathrm{\atom{C}{}{12}\atom{O}{}{}}}$ and $\mathcal{N}_{\mathrm{\atom{C}{}{13}\atom{O}{}{}}}$ are measured with the following equations by additionally assuming the abundance ratio of \atom{C}{}{12}\atom{O}{}{} to \atom{H}{}{}$_2$ $([\atom{C}{}{12}\atom{O}{}{}]/[\atom{H}{}{}_2])$ and \atom{C}{}{13}\atom{O}{}{} to \atom{C}{}{12}\atom{O}{}{} $([\atom{C}{}{13}\atom{O}{}{}]/[\atom{C}{}{12}\atom{O}{}{}])$:

\begin{eqnarray}
\left( \frac{\mathcal{N}_{\mathrm{^{12}CO}}}{\mathrm{cm^{-2}}} \right) =
\frac{[\atom{C}{}{12}\atom{O}{}{}]}{[\atom{H}{}{}_2]} & \times & \left( \frac{I_{\mathrm{^{12}CO}}}{\mathrm{K\> km\> s^{-1}}} \right) \nonumber \\
& \times & \left\{ \frac{X_{\mathrm{CO}}}{\mathrm{cm^{-2}\> (K\> km\> s^{-1})^{-1}}} \right\} \times \frac{1}{\eta_{\mathrm{f}}}
\end{eqnarray}

and
\begin{equation}
\left( \frac{\mathcal{N}_{\mathrm{^{13}CO}}}{\mathrm{cm^{-2}}} \right) =
\frac{[\atom{C}{}{13}\atom{O}{}{}]}{[\atom{C}{}{12}\atom{O}{}{}]} \times
\left( \frac{\mathcal{N}_{\mathrm{^{12}CO}}}{\mathrm{cm^{-2}}} \right),
\end{equation}
where $\eta_{\mathrm{f}}$ is a beam-filling factor.
Because the beam size of our \atom{C}{}{}\atom{O}{}{} data corresponds to 1.4$\>$kpc at NGC\,4303, this is quite a lot larger than the typical scale of molecular clouds.
As a consequence, observed integrated intensities are much less than the intrinsic intensity of a cloud.
Therefore, $\eta_{\mathrm{f}}$ is necessary for measuring a column density of a molecular cloud.
  
We first applied $X_{\mathrm{CO}} = 2.0 \times 10^{20}\> \mathrm{cm^{-2}\> (K\> km\> s^{-1})^{-1}}$ and assumed $[\atom{C}{}{12}\atom{O}{}{}]/[\atom{H}{}{}_2] = 8.5 \times 10^{-5}$ which is considered as the standard value for the Solar neighborhood (e.g., \cite{Frerking82}, \cite{Langer89}, \cite{Pineda08}).
\citet{Milam05} measured radial distribution of $[\atom{C}{}{12}]/[\atom{C}{}{13}]$ in the Galaxy and found that the relation $[\atom{C}{}{12}]/[\atom{C}{}{13}] = 6.21 D\> [\mathrm{kpc}] + 18.71$ where $D$ is the distance from the Galactic center.
$\atom{C}{}{13}$ is produced in the carbon-nitrogen-oxygen (CNO) cycle and this process occurs mainly in asymptotic giant branch (AGB) stars \citep{Pagel97}.
What causes the abundance gradient of $\atom{C}{}{13}$, however, is still unclear.
In this study, we adopted the local interstellar medium value of $[\atom{C}{}{12}\atom{O}{}{}]/[\atom{C}{}{13}\atom{O}{}{}]=68$ measured by \citet{Milam05}.

Since molecular clouds are not like discrete balls but a continuous medium, the line widths due to internal cloud motion $(dv)$ as an input parameter of RADEX was set depending on $\Delta V$.
In other words, they do not have the rigid boundary and internal cloud motion is likely to be affected by dynamics of surrounding gas.
If $\Delta V$ increases, $dv$ should also increase.
As mentioned above, the beam size of our observation is much larger than that of molecular clouds.
Therefore, one can assume that there are thousands of molecular clouds in the field of view of the telescope.
In addition, measured $\Delta V$ is a few tens km$\>$s$^{-1}$ (table 2), although the typical line width of a cloud is a few km s$^{-1}$.
Thus, $\Delta V$ represents the velocity dispersion among the molecular clouds in the beam.
We first defined $dv$ as 5.0 km s$^{-1}$ for the entire arms region.
This is the typical FWHM of emission lines from molecular clouds in the Galaxy (e.g., \cite{Heyer09}).
For other regions, we determined $dv$ by the following equation.
\begin{equation}
\left ( \frac{dv}{\mathrm{km\ s^{-1}}} \right)_i = 5.0\times \bigg( \frac{\Delta V}{\mathrm{km\ s^{-1}}} \bigg)_i \ \bigg/
\bigg( \frac{\Delta V}{\mathrm{km\ s^{-1}}} \bigg)_{\mathrm{entire\ arms}}
\end{equation}
where the subscript $i$ represents each region name and ${\Delta V}_{\mathrm{entire\ arms}} = 34\> \mathrm{km\ s^{-1}}$ (table \ref{tab:Table2}).

\subsubsection{How to estimate a beam-filling factor}
To set a beam-filling factor $\eta_{\mathrm{f}}$ in equation (4), we developed a method to estimate the best value of $\eta_{\mathrm{f}}$.
If the observed line is optically thick, molecular clouds do not overlap and sensitivity within the beam is uniform, $\eta_{\mathrm{f}}$ can be approximately assumed as
\begin{equation}
\eta_{\mathrm{f}} \simeq \frac{T_{\mathrm{peak,obs}}}{T_{\mathrm{model}}},
\end{equation}
where $T_{\mathrm{peak,obs}}$ is the observed peak intensity and $T_{\mathrm{model}}$ is the brightness temperature of a molecular cloud calculated by a model.

However, peak intensity is no longer available to estimate $\eta_{\mathrm{f}}$ for our observation.
Because the velocity resolution is 10 km s$^{-1}$ in our observation, it is not enough to measure the true peak intensity.
Moreover, profiles of observed spectra would be almost Dirac's delta function-like shape if molecular clouds in the beam do not have inter-clouds velocity dispersion.
Nevertheless, molecular clouds in the beam actually have inter-clouds velocity dispersion as mentioned above.
Hence, {$T_{\mathrm{peak,obs}}$} becomes less than $\eta_{\mathrm{f}} T_{\mathrm{model}}$.
The observed peak intensity does not reflect the amount of molecular clouds in the beam.

\begin{figure}[b!]
 \begin{center}
  \includegraphics[width=7.8cm]{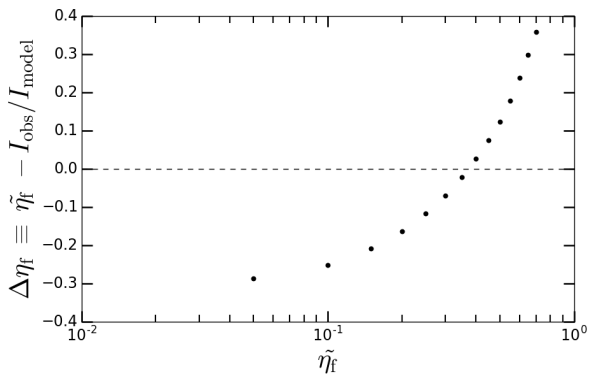}
 \end{center}
 \caption{$\Delta \eta_{\mathrm{f}}$ defined by equation (9) as a function of a given beam-filling factor $\tilde{\eta_{\mathrm{f}}}$. This is the result for the entire arms and $T_{\mathrm{kin}} = 15\>$K case. In this case, the optimum value of the beam-filling factor can be estimated to be 0.35--0.4.}
 \label{fig:Fig5}
\end{figure}

Even if the amount of molecular gas in the beam is constant, it is possible that the observed peak intensity changes due to inter-clouds velocity dispersion.
On the other hand, the observed integrated intensity would not depend on its velocity dispersion if the overlaps of molecular clouds in the beam can be ignored.
Therefore, we decided the best $\eta_{\mathrm{f}}$ using the observed integrated intensities in the following way.
As an analogous to equation (7) of the static-clouds motion case, $\eta_{\mathrm{f}}$ can be 
\begin{equation}
\eta_{\mathrm{f}} \simeq \frac{I_{\mathrm{obs}}}{I_{\mathrm{model}}},
\end{equation}
where $I_{\mathrm{obs}}$ is the observed integrated intensity and $I_{\mathrm{model}}$ is the integrated intensity calculated by RADEX.
Hence, when a given beam-filling factor $\tilde{\eta_{\mathrm{f}}}$ is the optimum value, it satisfies the following equation:
\begin{equation}
\Delta \eta_{\mathrm{f}} \equiv \tilde{\eta_{\mathrm{f}}} - \frac{I_{\mathrm{obs}}}{I_{\mathrm{model}}} = 0.
\end{equation}
Note that the fraction $I_{\mathrm{obs}}/I_{\mathrm{model}}$ in equations (8) and (9) is not constant; $I_{\mathrm{model}}$ is computed by RADEX and it needs a given beam-filling factor $\tilde{\eta_{\mathrm{f}}}$ as an input parameter according to equation (4).
Namely, $I_{\mathrm{model}}$ varies depending on a given parameter $\tilde{\eta_{\mathrm{f}}}$.

For the estimation of the optimum value of a beam-filling factor, \atom{C}{}{12}\atom{O}{}{}($J$\,=\,1--0) was used as the reference line, i.e., $I_{\mathrm{obs}}$ and $I_{\mathrm{model}}$ were applied to measured $I_{\mathrm{^{12}CO(1-0)}}$ with stacked spectra and results of calculations by RADEX, respectively.
Firstly, we input $\tilde{\eta_{\mathrm{f}}}=0.05$--0.7 with an interval of 0.05 and computed $\Delta \eta_{\mathrm{f}}$ with RADEX for each $\tilde{\eta_{\mathrm{f}}}$.
Figure \ref{fig:Fig5} shows how $\Delta \eta_{\mathrm{f}}$ is changed depending on each given $\tilde{\eta_{\mathrm{f}}}$ for the $T_{\mathrm{kin}} = 15\> \mathrm{K}$, entire arms case.
From this result, we can assume that the optimum value of the beam-filling factor ranges from 0.35 to 0.40, since the sign of $\Delta \eta_{\mathrm{f}}$ changes between these values.
Secondly, we input $\tilde{\eta_{\mathrm{f}}}$ from the value whose sign changes in the next step, to the next value of it with an interval of 0.01.
In figure \ref{fig:Fig5} case, for example, $\tilde{\eta_{\mathrm{f}}}$ is input in range of 0.35--0.40 with a step of 0.01.
Then we chose $\tilde{\eta_{\mathrm{f}}}$ whose $\Delta \eta_{\mathrm{f}}$ is the nearest to zero and considered it as the best $\eta_{\mathrm{f}}$.
In this way, the optimum value of the beam-filling factor is defined.

\begin{figure}[b!t!]
 \begin{center}
  \includegraphics[width=6.cm]{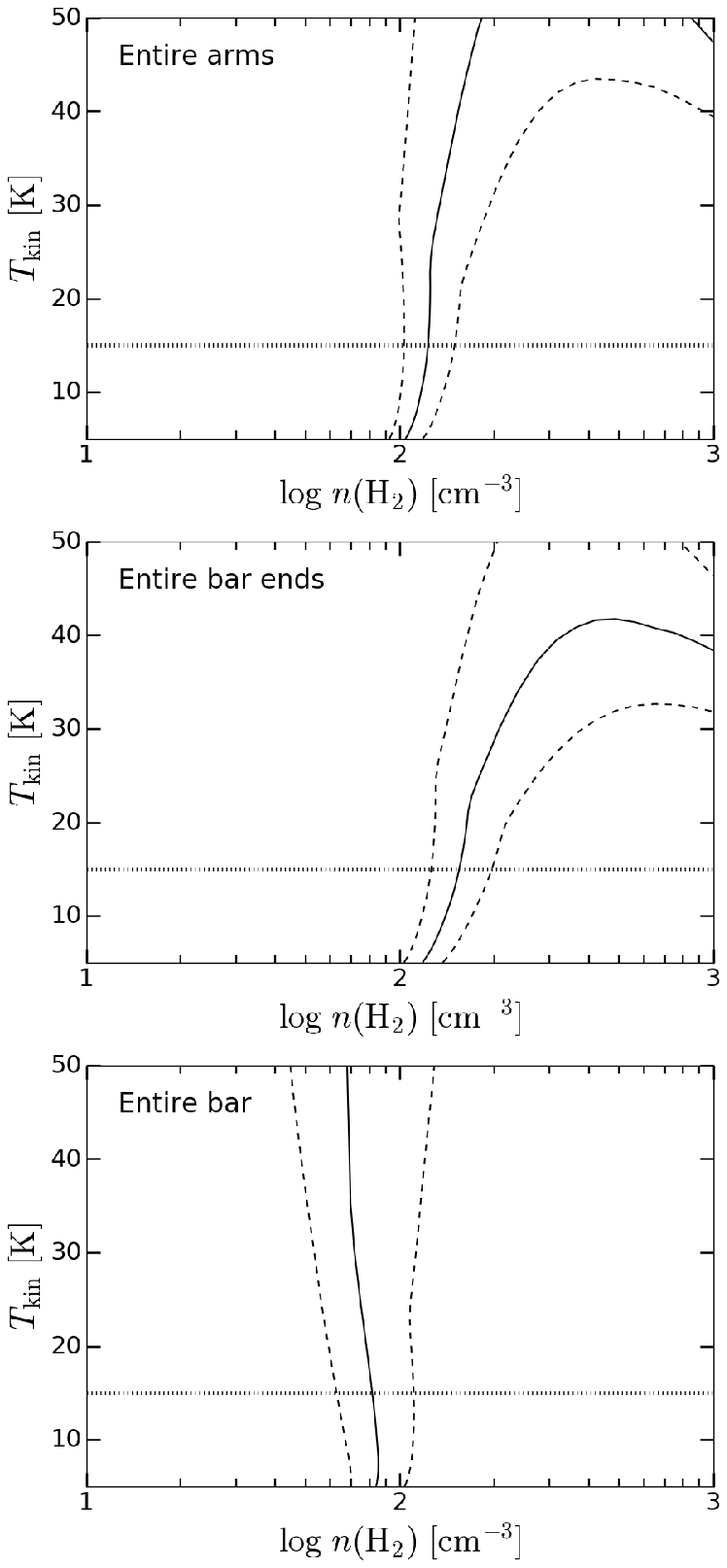}
 \end{center}
 \caption{Results of RADEX calculation for $T_{\mathrm{kin}} = 15\>$K in entire arms (top), entire bar ends (middle) and entire bar (bottom). Solid lines indicate the densities and the kinetic temperatures of molecular gas which reproduce observed $R_{12/13}$. Dashed lines represent 1\,$\sigma$ error of $R_{12/13}$. The intersections of the solid line and the dotted line correspond to the derived densities.}
 \label{fig:Fig6}
\end{figure}

\renewcommand\arraystretch{1.25}
\begin{table*}[b!t!h]
  \begin{center}
  \tbl{The derived molecular gas densities, the optimum values of the beam filling factor, the excitation temperatures of \atom{C}{}{12}\atom{O}{}{} and \atom{C}{}{13}\atom{O}{}{} in each region.}{
  \begin{tabular}{lcccccccccccc} \hline
  \tabcolsep = 5cm
                                           & \multicolumn{4}{c}{$T_{\mathrm{kin}} = 10\>\mathrm{K}$}          & \multicolumn{4}{c}{$T_{\mathrm{kin}} = 15\>\mathrm{K}$} & \multicolumn{4}{c}{$T_{\mathrm{kin}} = 20\>\mathrm{K}$} \\
 & $n(\rm{H_2})$ & $\eta_{\rm{f}}$ & $T_{\rm{ex,12}}$ & $T_{\rm{ex,13}}$ & $n(\rm{H_2})$ & $\eta_{\rm{f}}$ & $T_{\rm{ex,12}}$ & $T_{\rm{ex,13}}$ & $n(\rm{H_2})$ & $\eta_{\rm{f}}$ & $T_{\rm{ex,12}}$ & $T_{\rm{ex,13}}$ \\
 & [$10^2\>$cm$^{-3}$] &  & [K] & [K] & [$10^2\>$cm$^{-3}$] &  & [K] & [K] & [$10^2\>$cm$^{-3}$] &  & [K] & [K] \\ \hline
Northern bar & $1.3^{+0.7}_{-0.4}$ & 0.53 & 7.5 & 3.3 & $0.86^{+0.54}_{-0.27}$ & 0.38 & 9.3 & 3.4 & $0.68^{+0.45}_{-0.22}$ & 0.32 & 10 & 3.5 \\
Southern  bar & $1.3^{+0.6}_{-0.3}$ & 0.59 & 7.5 & 3.4 & $0.87^{+0.42}_{-0.24}$ & 0.42 & 9.3 & 3.5 & $0.68^{+0.35}_{-0.20}$ & 0.35 & 10 & 3.5 \\
$\>\>\>\>$Entire bar & $1.2^{+0.4}_{-0.3}$ & 0.56 & 7.5 & 3.3 & $0.82^{+0.29}_{-0.19}$ & 0.40 & 9.2 & 3.4 & $0.66^{+0.25}_{-0.16}$ & 0.34 & 10 & 3.5 \\
Northern bar end & $2.3^{+0.7}_{-0.5}$ & 0.54 & 8.5 & 3.8 & $1.5^{+0.5}_{-0.3}$ & 0.36 & 11 & 4.0 & $1.2^{+0.4}_{-0.3}$ & 0.29 & 13 & 4.2 \\
Southern bar end & $2.5^{+1.4}_{-0.7}$ & 0.69 & 8.5 & 3.9 & $1.7^{+1.0}_{-0.5}$ & 0.46 & 11 & 4.1 & $1.4^{+0.9}_{-0.4}$ & 0.37 & 13 & 4.3 \\
$\>\>\>\>$Entire bar ends & $2.3^{+0.6}_{-0.4}$ & 0.55 & 8.5 & 3.8 & $1.6^{+0.4}_{-0.3}$ & 0.37 & 11 & 4.0 & $1.3^{+0.4}_{-0.2}$ & 0.30 & 13 & 4.2 \\
Northern arm & $1.6^{+0.8}_{-0.5}$ & 0.61 & 8.0 & 3.5 & $1.1^{+0.6}_{-0.3}$ & 0.43 & 10 & 3.7 & $0.86^{+0.53}_{-0.27}$ & 0.35 & 12 & 3.7 \\
Eastern arm & $1.9^{+0.4}_{-0.3}$ & 0.53 & 8.2 & 3.6 & $1.2^{+0.3}_{-0.2}$ & 0.36 & 10 & 3.8 & $1.0^{+0.3}_{-0.2}$ & 0.30 & 12 & 3.9 \\
Western arm & $1.7^{+0.5}_{-0.3}$ & 0.57 & 8.1 & 3.6 & $1.1^{+0.4}_{-0.2}$ & 0.39 & 10 & 3.7 & $0.91^{+0.30}_{-0.20}$ & 0.32 & 12 & 3.8 \\
$\>\>\>\>$Entire arms & $1.9^{+0.4}_{-0.3}$ & 0.56 & 8.2 & 3.6 & $1.2^{+0.3}_{-0.2}$ & 0.37 & 10 & 3.8 & $1.0^{+0.2}_{-0.2}$ & 0.31 & 12 & 3.9 \\
  \hline
  \end{tabular}}
  \label{tab:Table3}
  \begin{tabnote}
  \end{tabnote}
  \end{center}
\end{table*}

As described in table \ref{tab:Table1}, NGC\,4303 has a nucleus classified as type H\,\emissiontype{II} \citep{Filippenko85c} and type 2 Seyfert \citep{Ho97d}.
Thus, it is considered that AGN feedback may have some impacts on molecular gas in the galactic center.
Moreover, chemical compositions and the $X_{\rm{CO}}$ in the central region may be different from the disk (e.g., \cite{Bolatto13}).
Therefore, we excluded the central region for non-LTE analysis.

\subsubsection{Results of non-LTE analysis}
Figure \ref{fig:Fig6} shows the calculation results of RADEX for the entire arms, entire bar ends and entire bar at the $T_{\mathrm{kin}}$ = 15$\>$K case.
Solid lines indicate $n(\mathrm{H_2})$ and $T_{\mathrm{kin}}$ which produce the measured $R_{12/13}$ in each region and dotted lines are its $1\, \sigma$ error.
The derived $n(\mathrm{H_2})$ corresponds to the intersection of solid and dotted lines.

All the derived densities of molecular gas and the optimum values of a beam-filling factor for each region and fixed kinetic temperature are listed in table \ref{tab:Table3}.
The molecular gas densities in the arms range from approximately 100--200$\>$cm$^{-3}$, which is almost consistent with the results of giant molecular clouds in the spiral arms of M\,51 \citep{Schinnerer10}.
In addition, our results fall into the range of strongly sub-thermal \atom{C}{}{13}\atom{O}{}{} as described in subsection 4.1.
Therefore, the assumption of fixed $T_{\rm{kin}}$ is not a serious problem.

According to table \ref{tab:Table3}, $n(\rm{H_2})$ in the bar is lower by approximately 37\% ($T_{\mathrm{kin}}$ = 10$\>$K), 31\% ($T_{\mathrm{kin}}$ = 15$\>$K), 34\% ($T_{\mathrm{kin}}$ = 20$\>$K) than that in spiral arms.
Hence, we confirmed that the density of molecular gas in the bar where the SFE also decreases is lower than that in the spiral arms. 
Moreover, we found that the $n({\mathrm{H_2}})$ in bar ends is larger by approximately 21\% ($T_{\mathrm{kin}} = 10\>$K), 33\% ($T_{\mathrm{kin}} = 15\>$K), 33\% ($T_{\mathrm{kin}} = 20\>$K) than that in the arms.
Calculated $\eta_{\mathrm{f}}$ are not so different for each region, whereas they are likely to depend on $T_{\rm{kin}}$ ($\eta_{\mathrm{f}}$ is $\sim 0.5$ for $T_{\mathrm{kin}} = 10\>$K, $\sim 0.4$ for $T_{\mathrm{kin}} = 15\>$K and $\sim 0.3$ for $T_{\mathrm{kin}} = 20\>$K).
These differences are caused by the fact that integrated intensities calculated by RADEX ($I_{\mathrm{model}}$) are sensitive to $T_{\mathrm{kin}}$ rather than $n(\mathrm{H_2})$.
The higher $T_{\mathrm{kin}}$ is, the larger $I_{\mathrm{model}}$ is in equations (8) and (9) and as a result, the ratio $I_{\mathrm{obs}} / I_{\mathrm{model}}$ becomes smaller and the optimum value of beam-filling factors is smaller.

The excitation temperatures of \atom{C}{}{12}\atom{O}{}{}($J$\,=\,1--0), $T_{\rm{ex,12}}$ and $T_{\rm{ex,13}}$ calculated by RADEX are also listed in table 3.
Since $\mathcal{N}_{\atom{C}{}{13}\atom{O}{}{}}$ is lower, \atom{C}{}{13}\atom{O}{}{} is strongly sub-thermalized. 
Optical depths of \atom{C}{}{12}\atom{O}{}{} and \atom{C}{}{13}\atom{O}{}{}, $\tau_{12}$ and $\tau_{13}$ is almost constant among the regions: $\tau_{12}$ = 20, 19--20, 18--20, and $\tau_{13}$ = 0.77--0.82, 1.0--1.1, 1.2--1.3 for the $T_{\rm{kin}}$ = 10, 15, 20$\>$K cases, respectively.
One may assume that the optical depths could be lower in the bar than that of others, because the line widths are larger.
We note that the large column densities in the bar cancel this effect [see also equation (21) in \cite{vanderTak07}].
In addition, the derived $\tau_{13}$ are much larger than those in \citet{Cormier18}, although the results of $R_{12/13}$ for NGC\,4303 are consistent with theirs.
This may be because a $T_{\rm{ex,13}}$ they assumed is much higher (20--30$\>$K) than our derived ones, which lead to lower value of $\tau_{13}$ [see also equation (3) in \cite{Cormier18}].
Several studies have reported $\tau_{13} \sim 1$, for example, \citet{RD10}, \citet{Hacar16} and \citet{Torii18}.
Thus, our results of $\tau_{13} \sim 1$ are not unique.

\subsection{The cause of SFE and molecular gas density variations}
\begin{figure}[b!t!p!]
 \begin{center}
  \includegraphics[width=5.9cm]{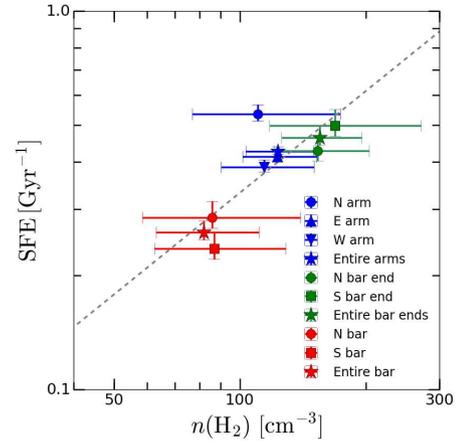}
 \end{center}
 \caption{Correlation between SFE and molecular gas density $n(\mathrm{H_2})$ derived from RADEX for the $T_{\mathrm{kin}} = 15\>$K case. The gray dashed line indicates the least square fitting.}
 \label{fig:Fig7}
\end{figure}

According to \citet{Leroy08}, it can be considered that SFE is proportional to the reciprocal of the time scale of star formation $\tau_{\mathrm{SF}}$; $\mathrm{SFE} \propto {\tau_{\mathrm{SF}}}^{-1}$.
If $\tau_{\rm{SF}}$ is determined by free-fall time, $\tau_{\mathrm{SF}} \propto 1/\sqrt{n(\mathrm{H_2})}$ and thus, SFE can satisfy $\mathrm{SFE} \propto \sqrt{n(\mathrm{H_2})}$.
Therefore, we examine the correlation between regionally averaged SFEs measured in subsection 3.3 and the square root of molecular gas densities $\sqrt{n(\mathrm{H_2})}$.
We found that the correlation coefficient is 0.78, 0.78, 0.79 for $T_{\mathrm{kin}} =\ $10, 15, 20$\>$K, respectively.
It is clear that there is a positive correlation between SFEs and $\sqrt{n(\mathrm{H_2})}$ among the galactic regions.
We also show the double logarithmic correlation plot between SFEs and $n(\mathrm{H_2})$ for the $T_{\mathrm{kin}} = 15\>$K case in figure \ref{fig:Fig7}.
The gray dashed line in the figure indicates the least squares law fitting.
In the double logarithmic plot case, the correlation coefficient is 0.83, 0.82, 0.83 for $T_{\mathrm{kin}} = \ $10, 15, 20$\>$K, respectively.
This positive correlation is the second case subsequently to \citet{Muraoka16b} for NGC\,2903.
According to \citet{Muraoka16b}, the correlation coefficient for the double logarithmic SFE--$n(\mathrm{H_2})$ plot is 0.71, which is the similar result to that for this study (note that they reported not the correlation coefficient but the coefficient of determination $R^2$).
To investigate whether the relation between SFE and $n(\rm{H_2})$ is almost the same for other galaxies, more samples and further studies are needed.

The northern arm seems to have a deviation from the correlation as shown in figure \ref{fig:Fig7}.
The reason may be that molecular gas has been halfway to depletion although it was plenty before the current active star formation in the northern arm.
As a result, the SFE becomes relatively larger than that of others.
It is also possible that an interaction with other galaxies may produce compact dense molecular clouds in the northern arm, which could not be detected in our observation, even though the kpc-scale averaged density of molecular gas is lower.
The deviation from the usual spider pattern for the mean velocity field is seen at near the root of the northern arm according to \ref{fig:Fig1} (c).
In fact, \citet{Cayatte90} argued that many supernovae and H\,\emissiontype{II} regions in NGC\,4303 indicate recent star-burst activity, and this may be due to a tidal interaction with its companions NGC\,4292 and NGC\,4301 (NGC\,4303A). 

Even though there are some deviations, we can conclude that the variation of SFEs within NGC\,4303 is caused by the different densities of molecular gas.
Then, the next question appears: why do the densities of molecular gas vary among the galactic structures?
We discuss the reason for the various molecular gas densities in the following.

Many previous studies based on both observations and simulations have reported that non-circular motion is large in the bar (e.g., \cite{Athanassoula92}, \cite{Raynaud98}, \cite{Watanabe11}, \cite{Fujimoto14}, \cite{Maeda18}, \cite{Sun18}).
Therefore, we investigate the reason of the lower density condition of molecular gas in the bar by focusing on dynamics of molecular gas in each region.
As discussed in subsection 4.1, the FWHM of observed and stacked spectra $\Delta V$ represents the velocity dispersion of inter-molecular clouds in the beam.
Since the galactic disk is tilted against the observer, the observed velocity becomes larger by a factor of $\sin i$ than that for the face-on case. 
Hence, the quantity $\Delta V/ \sin i$ could indicate the velocity dispersion of inter-molecular clouds projected on the galactic disk.
We examine the relation between $n(\mathrm{H_2})$ and $\Delta V/ \sin i$.

\begin{figure}[b!t!]
 \begin{center}
  \includegraphics[width=7.9cm]{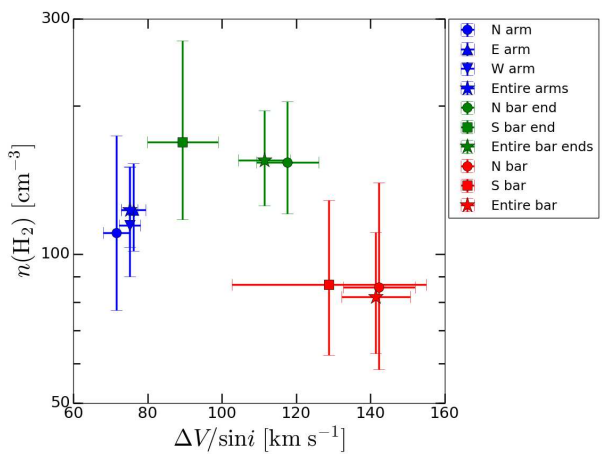}
 \end{center}
 \caption{Relationship between derived $n(\mathrm{H_2})$ and $\Delta V/\sin i$ for the $T_{\rm{kin}}=15\>$K case.}
 \label{fig:Fig8}
\end{figure}

Figure \ref{fig:Fig8} shows $n(\mathrm{H_2})$ in the $T_{\rm{kin}} = 15\>$K case against $\Delta V / \sin i$ for each region.
It is clear that when $\Delta V/ \sin i \lesssim 100\>\mathrm{km\> s^{-1}}$ (from the arms to the bar end region), $n(\mathrm{H_2})$ increases as $\Delta V/\sin i$ increases.
On the other hand, when $\Delta V/\sin i \gtrsim 100\>\mathrm{km\>s^{-1}}$ (from the bar end to the bar region), $n(\mathrm{H_2})$ decreases as $\Delta V/\sin i$ increases.
That is, we found that if inter-clouds velocity dispersion is moderate, density of molecular gas get denser and if velocity dispersion is too large, it becomes lower inversely: the $n(\mathrm{H_2})$--$\Delta V /\sin i$ relationship with the asymmetric-``$\Lambda$" like shape.
This can be also seen in $T_{\rm{kin}}=10\>$K and 20$\>$K cases.
In summary, we can conclude that density of molecular gas could vary depending on the dynamical effect such as inter-clouds velocity dispersion.
From the next paragraph, we further discuss what processes influence on $n(\rm{H_2})$ for the lower and higher inter-clouds velocity dispersion regimes, respectively.

For the $\Delta V/ \sin i \lesssim 100\> \mathrm{km\> s^{-1}}$ (the positive correlation regime between $n(\mathrm{H_2})$ and $\Delta V/\sin i$) case, moderate velocity dispersion seems to enhance density of molecular gas.
The tendency is explained if cloud-cloud collisions \citep{Habe92} occur more frequently due to reasonable inter-clouds' motion, which would increase the probability of collisions or interactions with molecular clouds.

In figure \ref{fig:Fig8}, molecular gas density is increased in bar ends where the inter-clouds velocity dispersions are moderate.
It is possible that orbits of molecular gas have self-intersections around the bar end (e.g., \cite{Blitz93}, \cite{Downes96}).
This fact may be the reason for moderate velocity dispersion and the increased probability of collisions.
In consequence of such processes, the molecular gas densities would become higher in the bar ends.
In fact, for the star-forming region W\,43, which is considered to be located in the bar end of the Galaxy (e.g., \cite{Nguyen11}), it would be a cloud-cloud collisions site (e.g., \cite{Motte14}).

On the other hand, for the $\Delta V/\sin i \gtrsim 100\> \mathrm{km\> s^{-1}}$ (the negative correlation) regime, collision speed could become higher than that of the moderate case.
Therefore, because of such high collision speed, molecular clouds could be destroyed (e.g., \cite{Tubbes82}).
In another aspect of destructive colliding, \citet{Takahira14} showed that if molecular clouds collide at high speed, many molecular cloud cores are formed.
However, they do not have enough time to accrete molecular gas.
As a result, molecular cloud cores cannot grow and mean density does not become higher.
In addition to the cloud-cloud collision scenario, large velocity dispersion of inter-clouds should enhance inner-cloud velocity dispersion, or turbulence.
This causes molecular clouds to gravitationally contract hardly (e.g., \cite{Sorai12}, \cite{Nimori13}).
Finally, strong shear motion could become dominant in such a vigorous environment. 
Therefore, molecular clouds would be stripped, shredded and finally destroyed (e.g., \cite{Athanassoula92}).
Understanding which is the most dominant process in the bar is a remaining issue.

\section{Conclusions}
We present the results of simultaneous \atom{C}{}{12}\atom{O}{}{}($J$\,=\,1--0) and \atom{C}{}{13}\atom{O}{}{}($J$\,=\,1--0) mappings toward the nearby barred spiral galaxy NGC\,4303 with the Nobeyama 45-m telescope and the receiver FOREST as the NRO legacy project COMING.
Main conclusions in this paper are as follows:

\begin{enumerate}
\renewcommand{\labelenumi}{(\arabic{enumi})}
\item Galactic structures can be seen in \atom{C}{}{12}\atom{O}{}{}($J$\,=\,1--0).
\atom{C}{}{13}\atom{O}{}{}($J$\,=\,1--0) is detected mainly at the galactic center and \atom{C}{}{}\atom{O}{}{18}($J$\,=\,1--0) is not detected from our map.

\item We separated NGC\,4303 into eight regions (the center, northern bar, southern bar, northern bar end, southern bar end, northern arm, eastern arm, western arm) based on its structures.
Regionally averaged molecular gas surface densities ($\Sigma_{\mathrm{mol}}$), surface densities of star-formation rates ($\Sigma_{\mathrm{SFR}}$) and star-formation efficiencies (SFE) are measured in each region.

\item The SFE in the bar is lower by approximately 39\% than that in the spiral arms.
This tendency is consistent with previous studies about barred spiral galaxies.
In addition, the SFE in the bar ends is slightly larger ($\sim 10\%$) than that in the spiral arms.

\item From stacking analysis for each galactic region, we can successfully detect \atom{C}{}{12}\atom{O}{}{}($J$\,=\,1--0) and \atom{C}{}{13}\atom{O}{}{}($J$\,=\,1--0) lines and measure their integrated intensities for all regions.
\atom{C}{}{}\atom{O}{}{18}($J$\,=\,1--0) line cannot be detected significantly even with stacking.

\item Based on measured integrated intensity ratios of \atom{C}{}{12}\atom{O}{}{}($J$\,=\,1--0) and \atom{C}{}{13}\atom{O}{}{}($J$\,=\,1--0) line from stacked spectra, we derive volume densities of molecular gas $n(\mathrm{H_2})$ at a fixed kinetic temperature $T_{\mathrm{kin}}$ for each region via non-LTE analysis.
To perform non-LTE analysis, we developed a way to obtain the correct beam dilution effect.
As a result, beam-filling factors are not so different among the regions.

\item Results of non-LTE analysis show the $n(\rm{H_2})$ in the bar is lower by 31--37\% than that in the arms.
We confirm that in the bar where the SFE is low, volume density of molecular gas is also low.
Moreover, the $n(\rm{H_2})$ in the bar ends is higher by 21--33\% than that in the arms.

\item There is a positive correlation between SFE and derived $n(\mathrm{H_2})$ among the regions.
The correlation coefficient is 0.82--0.83.
Hence, the variations of SFE within NGC\,4303 would be caused by the different volume densities of molecular gas.

\item We found that when inter-clouds velocity dispersion is moderate ($\Delta V/\sin i \lesssim 100\> \mathrm{km\ s^{-1}}$), $n(\mathrm{H_2})$ tends to be larger as velocity dispersion increases.
On the other hand, when velocity dispersion is too large ($\Delta V/\sin i \gtrsim 100 \>\rm{km\>s^{-1}}$), $n(\mathrm{H_2})$ decreases as velocity dispersion increases.
This suggests that volume density of molecular gas would be influenced by dynamical effects indicated by inter-clouds velocity dispersion.

\item The relation between $n(\rm{H_2})$ and $\Delta V/\sin i$ could be interpreted as follows:
In the moderate $\Delta V/\sin i$ situation, cloud-cloud collisions could occur frequently.
By contrast, when $\Delta V/\sin i$ exceeds 100$\>$km$\ $s$^{-1}$, molecular clouds would be shredded and destroyed.
It is also possible that molecular clouds are hard to shrink by self-gravity due to enhanced inner turbulences and cores in molecular clouds being unable to grow enough because of high collision speed.
\end{enumerate}

\section*{Acknowledgments}
We are grateful to NRO staff for setting up and operating the Nobeyama 45-m telescope system, receivers and other equipments, and for supporting our project.
We are also grateful to an anonymous referee for the careful reading of our manuscript and the constructive comments that significantly improved this paper.
The Nobeyama 45-m radio telescope is operated by Nobeyama Radio Observatory, a branch of National Astronomical Observatory of Japan.
This research has made use of the NASA/IPAC Extragalactic Database, which is operated by the Jet Propulsion Laboratory, California Institute of Technology, under contract with the National Aeronautics and Space Administration.


\begin{thebibliography}{99}
\bibitem[Aalto et al.(1991)]{Aalto91} Aalto, S., Black, J. H., Johansson, L. E. B., \& Booth, R. S. 1991, \aap, 249, 323
\bibitem[Aalto et al.(1995)]{Aalto95} Aalto, S., Booth, R. S., Black, J. H., \& Johansson, L. E. B. 1995, \aap, 300, 369
\bibitem[Argyle \& Eldridge(1990)]{Argyle90} Argyle, R. W., \& Eldridge, P. 1990, \mnras, 243, 504
\bibitem[Athanassoula (1992)]{Athanassoula92} Athanssoula, E. 1992, \mnras, 259, 345
\bibitem[Bendo et al.(2012)]{Bendo12} Bendo, G. J., Galliano, F., \& Madden, S. C. 2012, \mnras, 423, 197
\bibitem[Bigiel et al.(2016)]{Bigiel16} Bigiel, F., et. al. 2016, \apj, 822, L26
\bibitem[Blitz et al.(1993)]{Blitz93}Blitz, L., Binney, J., Lo, K. Y., Bally, J., \& Ho, P. T. P. 1993, \nat, 361, 417
\bibitem[Bolatto et al.(2013)]{Bolatto13} Bolatto, D. A., Wolfire, M., \& Leroy, A. K. 2013, \araa, 51, 207 
\bibitem[Calzetti et al.(2007)]{Calzetti07} Calzetti, D., et al. 2007, \apj, 666, 870
\bibitem[Cao et al.(2017)]{Cao17} Cao, Y., Wong, T., Xue, R., Bolatto, A. D., Blitz, L., Vogel, S. N., Leroy, A. K., \& Rosolowsky, E. 2017, \apj, 847, 33 
\bibitem[Cayatte et al.(1990)]{Cayatte90} Cayatte, V., van Gorkom, J. H., Balkowski, C., \& Kotanyi, C. 1990, \aj, 100, 604
\bibitem[Cormier et al.(2018)]{Cormier18} Cormier, D., et al. 2018, \mnras, 475, 3909
\bibitem[Davis(2014)]{Davis14} Davis, T. A. 2014, \mnras, 445, 2378
\bibitem{} de Vaucouleurs, G., de Vaucouleurs, A., Corwin, H. G., Buta, R. J., Paturel, G., \& Fouque, P. 1991, Third Reference Catalogue of Bright Galaxies (RC3) (New York: Springer-Verlag)
\bibitem[Downes et al.(1996)]{Downes96} Downes, D., Reynaud, D., Solomon, P. M., \& Radford, S. J. E. 1996, \apj, 461, 186
\bibitem[Emerson \& Gr\"ave(1988)]{Emerson88} Emerson, D. T., \& Gr\"ave, R. 1988, \aap, 190, 353
\bibitem[Filippenko \& Sargent (1985)]{Filippenko85c} Filippenko, A. V., \& Sargent, W. L. 1985, \apj, 57, 503
\bibitem[Frerking et al.(1982)]{Frerking82} Frerking, M, A., Langer, W, D., \& Wilson, R, W. 1982, \apj, 262, 590
\bibitem[Fujimoto et al.(2014)]{Fujimoto14} Fujimoto, Y., Tasker, E. J., Wakayama, M., \& Habe, A. 2014, \mnras, 439, 936
\bibitem[Gao \& Solomon(2004)]{Gao04} Gao, Y., \& Solomon, P. M. 2004, \apjs, 152, 63
\bibitem[Gao et al.(2007)]{Gao07} Gao, Y., Carilli, C. L., Solomon, P. M., \& Vanden Bout P. A. 2007, \apj, 660, L93
\bibitem[Gil de Paz et al. (2007)]{Paz07} Gil de Paz, A., et al. 2007, \apjs, 173, 185
\bibitem[Goldreich \& Kwan(1974)]{Goldreich74} Goldreich, P., \& Kwan, J. 1974, \apj, 189, 441
\bibitem[Habe \& Ohta(1992)]{Habe92} Habe, A., \& Ohta, K. 1992, \pasj, 44, 203
\bibitem[Hacar et al.(2016)]{Hacar16} Hacar, A., Alves, J., Burkert, A., \& Goldsmith, P. 2016, \aap, 591, A104 
\bibitem[Handa et al.(1991)]{Handa91} Handa, T., Sofue, Y., \& Nakai, N. 1991, in IAU Symp. 146, Dynamics of Galaxies and Their Molecular Cloud Distributions, ed. F. Combes \& F. Casoli (Dordrecht: Kluwer Academic Publishers), 156
\bibitem[Heyer et al.(2009)]{Heyer09} Heyer, M., Krawczyk, C., Duval, J., \& Jackson, L. M. 2009, \apj, 699, 1092
\bibitem[Hirota et al.(2010)]{Hirota10} Hirota, A., Kuno, N., Sato, N., Nakanishi, H., Tosaki, T., \& Sorai, K. 2010, \pasj, 62, 1261
\bibitem[Hirota et al.(2014)]{Hirota14} Hirota, A., et al. 2014, \pasj, 66, 46
\bibitem[Ho, Filippenko \& Sargent (1997)]{Ho97d} Ho, L. C., Filippenko, A. V., \& Sargent, W. L. 1997, \apjs, 112, 315
\bibitem[Iono et al.(2007)]{Iono07} Iono, D., et al. 2015, \apj, 659, 283
\bibitem[Jim{\'e}nez-Donaire et al.(2017)]{JD17} Jim{\'e}nez-Donaire, M. J., et al. 2017, \apjl, 836, L29 
\bibitem[Kamazaki et al.(2012)]{Kamazaki12} Kamazaki, T., et. al. 2012, \pasj, 64, 29
\bibitem[Kennicutt et al.(2007)]{Kennicutt07} Kennicutt, R. C., Jr., et al. 2007, \apj, 671, 333 
\bibitem[Koda \& Sofue(2006)]{Koda06} Koda, J., \& Sofue, Y. 2006, \pasj, 58, 299
\bibitem[Kroupa(2001)]{Kroupa01} Kroupa, P. 2001, \mnras, 322, 231
\bibitem[Kuno et al.(2007)]{Kuno07} Kuno, N., et al. 2007, \pasj, 59, 117
\bibitem[Kuno et al.(2011)]{Kuno11} Kuno, N., et. al. 2011, General Assembly and Scientific Symposium, XXXth URSI, JP2-19
\bibitem[Langer et al.(1984)]{Langer84} Langer, W. D., Graedel, T. E., Frerking, M.~A., \& Armentrout, P. B. 1984, \apj, 277, 581
\bibitem[Langer et al.(1989)]{Langer89} Langer, W. D., Wilson, R, W., Goldsmith, P, F., \& Beichman, C, A. 1989, \apj, 337, 355
\bibitem[Leroy et al.(2008)]{Leroy08} Leroy, A. K., Walter, F., Brinks, E., Bigiel, F., de Blok, W. J. G., Madore, B., \& Thornley, M. D. 2008, \apj, 136, 2782
\bibitem[Liszt(2007)]{Liszt07} Liszt, H. S. 2007, \aap, 476, 291 
\bibitem[Maeda et al.(2018)]{Maeda18} Maeda, F., Ohta, K., Fujimoto, Y., Habe, A., \& Baba, J. 2018, \pasj, 70, 37
\bibitem[Mei et al.(2007)]{Mei07} Mei, S., et al. 2007, \apj, 655, 144
\bibitem[Meier \& Turner(2004)]{Meier04} Meier, D. S., \& Turner, J. L. 2004, \aj, 127, 2069
\bibitem[Meier, Turner \& Hurt(2008)]{Meier08} Meier, D. S., Turner, J. L., \& Hurt, R. L. 2008, \apj, 675, 281
\bibitem[Milam et al.(2005)]{Milam05} Milam, S. N., Savage, C., Brewstar, M. A., Ziurys, L. M., \& Wyckoff, S. 2005, \apj, 634, 1126
\bibitem[Minamidani et al.(2011)]{Minamidani11} Minamidani, T., et al. 2011, \aj, 141, 73
\bibitem[Minamidani et al.(2016)]{Minamidani16} Minamidani, T., et al. 2016, Proc. Spire, 9914, 99141Z
\bibitem[Momose et al.(2010)]{Momose10} Momose, R., Okumura, S. K., Koda, J., \& Sawada, T. 2010, \apj, 721, 383
\bibitem[Morokuma-Matsui et al.(2015)]{Morokuma15} Morokuma-Matsui, K., Sorai, K., Watanabe, Y., \& Kuno, N. 2015, \pasj, 67, 2
\bibitem[Motte et al.(2014)]{Motte14} Motte, F., et. al. \aap, 2014, 571, A32
\bibitem[Muraoka et al. (2009)]{Muraoka09} Muraoka, K., et al. 2009, \pasj, 61, 163
\bibitem[Muraoka et al.(2012)]{Muraoka12} Muraoka, K., Tosaki, T., Miura, R., Onodera, S., Kuno, N., Nakanishi, K., Kaneko, H., \& Komugi, S. 2012, \pasj, 64, 3
\bibitem[Muraoka et al.(2016)]{Muraoka16b} Muraoka, K., et al. 2016, \pasj, 68, 89
\bibitem[Nguyen et al.(2011)]{Nguyen11} Nguyen Q. L., et al. 2011, \aap, 529, A41
\bibitem[Nimori et al.(2013)]{Nimori13} Nimori, M., Habe, A., Sorai, K., Watanabe, Y., Hirota, A., \& Namekata, D. 2013, \mnras, 429, 2175
\bibitem[Pagel (1997)]{Pagel97} Pagel, B. 1997, Nucleosynthesis and Chemical Evolution of Galaxies (Cambridge: Cambridge Univ. Press)
\bibitem[Paglione et al.(2001)]{Paglione01} Paglione, T. A. D., et al. 2001, \apjs, 135, 183
\bibitem[Pan et al.(2015)]{Pan15} Pan, H.-A., Kuno, N., Sorai, K., \& Umei, M. 2015, \pasj, 67, 116
\bibitem[Pineda et al.(2008)]{Pineda08} Pineda, J. E., Caselli, P., \& Goodman, A. A. 2008, \apj, 679, 481
\bibitem[Raynaud \& Downes(1998)]{Raynaud98} Reynaud, D., \& Downes, D. 1998, \aap, 337, 671
\bibitem[Roman-Duval et al.(2010)]{RD10} Roman-Duval, J., Jackson, J. M., Heyer, M., Rathborne, J., \& Simon, R. 2010, \apj, 723, 492
\bibitem[Salo et al.(2015)]{Salo15} Salo, H., et al. 2015, \apjs, 219, 4
\bibitem[Salpeter(1955)]{Salpeter55} Salpeter, E. E. 1955, \apj, 121, 161
\bibitem[Sawada et al.(2008)]{Sawada08} Sawada, T., et al. 2008, \pasj, 60, 445
\bibitem[Schinnerer et al.(2010)]{Schinnerer10} Schinnerer, E., Wei\ss, A., Aalto, S., \& Scoville, N. Z. 2010, \apj, 719, 1588
\bibitem[Schruba et al.(2011)]{Schruba11} Schruba, A., et al. 2011, \aj, 142, 37
\bibitem[Scoville \& Solomon(1974)]{Scoville74} Scoville, N. Z., \& Solomon, P. M. 1974, \apj, 187, L67
\bibitem[Sheth et al.(2000)]{Sheth00} Sheth, K., Regan, M. W., Vogel, S. N., \& Teuben, P. J. 2000, \apj, 532, 221
\bibitem[Sheth et al.(2002)]{Sheth02} Sheth, K., Vogel, S. N., Regan, M. W., Teuben, P. J., Harris, A. I., \& Thornley, M. D. 2002, \aj, 124, 2581
\bibitem[Sheth et al. (2010)]{Sheth10} Sheth, K., et al. 2010, \pasp, 122, 1397 
\bibitem[Sobolev (1960)]{Sobolev60} Sobolev, V. 1960, Moving envelopes of stars (Harvard University Press)
\bibitem[Solomon et al.(1992)]{Solomon92} Solomon, P. M., Downes, D., \& Radford, S. J. E. 1992, \apj, 387, L55
\bibitem[Sorai et al.(2012)]{Sorai12} Sorai, K., et al. 2012, \pasj, 64, 51
\bibitem{} Sorai, K., et al. 2019, \pasj, submitted
\bibitem[Sun et al.(2018)]{Sun18} Sun, J., et. al. 2018, \apj, 860, 172
\bibitem[Sz{\H u}cs et al.(2014)]{Szucs14} Sz{\H u}cs, L., Glover, S. C. O., \& Klessen, R. S. 2014, \mnras, 445, 4055
\bibitem[Takahira et al.(2014)]{Takahira14} Takahira, K., Tasker, E. J., \& Habe, A. 2014, \apj, 792, 63
\bibitem[Torii et al.(2018)]{Torii18} Torii, K., et al. 2018, \pasj, preprint (doi: 10.1093/pasj/psy098)
\bibitem[Tubbes(1982)]{Tubbes82} Tubbs, A. D. 1982, \apj, 255, 458
\bibitem[Usero et al.(2015)]{Usero15} Usero, A., et al. 2015, \apj, 150, 115
\bibitem[van der Tak et al.(2007)]{vanderTak07} van der Tak, F. F. S., Black, J. H., Sch\"oier, F. L., Jansen, D. J., \& van Dishoeck, E. F. 2007, \aap, 468, 627
\bibitem[Watanabe et al.(2011)]{Watanabe11} Watanabe, Y., Sorai, K., Kuno, N., \& Habe, A. 2011, \mnras, 411, 1409

\end{thebibliography}
\end{document}